\documentclass{ws-ijmpd}
\usepackage[super,compress]{cite}
\usepackage{bbding,enumitem}

\usepackage{pifont}%

\usepackage{subfigure}
\usepackage{multirow}
\usepackage{cancel}
\usepackage{array}

\RequirePackage[colorlinks,citecolor=blue,urlcolor=blue,linkcolor=blue]{hyperref}

\usepackage{amsmath}
\usepackage{amssymb}
\newcommand{\be}{\begin{equation}}
\newcommand{\ee}{\end{equation}}

\AtBeginDocument{}

\usepackage[normalem]{ulem} 

\begin{document}

\markboth{{\bf Reginald Christian Bernardo and Kin-Wang Ng}}
{{\bf Charting the Nanohertz Gravitational Wave Sky with Pulsar Timing Arrays}}

%
\catchline{}{}{}{}{}
%

\title{
CHARTING THE NANOHERTZ GRAVITATIONAL WAVE SKY \\ WITH PULSAR TIMING ARRAYS
}

\author{{\bf Reginald Christian Bernardo}
}

\address{Asia Pacific Center for Theoretical Physics\\
Pohang 37673, Korea\\
reginald.bernardo@apctp.org}

\author{{\bf Kin-Wang Ng}
}

\address{Institute of Physics, Academia Sinica\\
Taipei 11529, Taiwan\\
nkw@phys.sinica.edu.tw}

\address{Institute of Astronomy and Astrophysics, Academia Sinica\\
Taipei 11529, Taiwan}

\maketitle


\begin{abstract}
In the summer of 2023, the pulsar timing arrays (PTAs) announced a compelling evidence for the existence of a nanohertz stochastic gravitational wave background (SGWB). Despite this breakthrough, however, several critical questions remain unanswered: What is the source of the signal? How can cosmic variance be accounted for? To what extent can we constrain nanohertz gravity? When will individual supermassive black hole binaries become observable? And how can we achieve a stronger detection? These open questions have spurred significant interests in PTA science, making this an opportune moment to revisit the astronomical and theoretical foundations of the field, as well as the data analysis techniques employed. In this review, we focus on the theoretical aspects of the SGWB as detected by PTAs. We provide a comprehensive derivation of the expected signal and its correlation, presented in a pedagogical manner, while also addressing current constraints. Looking ahead, we explore future milestones in the field, with detailed discussions on emerging theoretical considerations such as cosmic variance, the cumulants of the one- and two-point functions, subluminal gravitational waves, and the anisotropy and polarization of the SGWB.
\end{abstract}



\section{Introduction}
\label{sec:introduction}

The discovery of gravitational waves (GWs) marked a pivotal moment in our understanding of the Universe, opening a new era in astronomy and deepening our grasp of gravity's most extreme manifestations. From the first direct detection, which captured ripples in spacetime from merging black holes, to the observation of hundreds of such events from Solar mass compact objects, each detection has provided invaluable insights into the strong gravity regime \cite{LIGOScientific:2016aoc, KAGRA:2021vkt, LIGOScientific:2021sio}. Yet, the quest to fully explore the GW spectrum is far from over. The next major breakthrough on the horizon is the detection of a stochastic gravitational wave background (SGWB) \cite{Caprini:2018mtu, Christensen:2018iqi, Romano:2019yrj, Moore:2021ibq, NANOGrav:2020spf, Pol:2022sjn, Staelens:2023xjn}---hums of spacetime arising from a superposition of countless unresolved sources. While efforts to detect this background in the sub-kilohertz and millihertz bands continue with ground-based \cite{LIGOScientific:2009qal, Shoemaker:2019bqt, Lehoucq:2023zlt} and future space-based observatories \cite{LISACosmologyWorkingGroup:2022kbp, Cheng:2022vct, Liang:2021bde, Muratore:2023gxh}, recent astronomical advances have shifted attention to the nanohertz regime.

Pulsar timing arrays (PTAs) have emerged as a powerful tool, offering compelling evidence for the existence of a nanohertz SGWB \cite{NANOGrav:2023gor, Reardon:2023gzh, EPTA:2023fyk, Xu:2023wog}. PTAs monitor the precise timing of millisecond pulsars across our Galaxy, and recent observations have revealed a correlated common-spectrum process consistent with a SGWB. The SGWB signal, characterized by the quadrupolar pattern predicted by the Hellings and Downs (HD) curve \cite{Hellings:1983fr}, is traditionally interpreted as originating from a population of supermassive black hole binaries (SMBHB) \cite{Sazhin:1978myk, Detweiler:1979wn, Phinney:2001di, Wyithe:2002ep, Sesana:2004sp, Sesana:2008mz, Vigeland:2016nmm, Burke-Spolaor:2018bvk, Liu:2021ytq, Sato-Polito:2023gym, Sato-Polito:2023spo, Bi:2023tib, Sato-Polito:2024lew, Sah:2024oyg, Raidal:2024odr, Sah:2024etc}. However, the spectral properties of the detected signal, which appear bluer than expected compared to the SMBHB model, have sparked a tantalizing debate about its origin \cite{Chen:2019xse, Ellis:2020ena, Vagnozzi:2020gtf, Benetti:2021uea, NANOGrav:2021flc, Buchmuller:2021mbb, Xue:2021gyq, Sharma:2021rot, EPTA:2023xxk, Vagnozzi:2023lwo, Figueroa:2023zhu, Ellis:2023dgf, Saeedzadeh:2023biq, Liu:2023pau, Liu:2023hpw, Chen:2023bms, Liu:2023ymk, Jin:2023wri, Huang:2023chx, Ye:2023tpz, Wang:2023sij, Wang:2023ost, Zhu:2023lbf, Jiang:2023gfe, Bian:2023dnv, Jiang:2024dxj, Winkler:2024olr, Agazie:2024kdi, Calza:2024qxn, Papanikolaou:2024fzf, Papanikolaou:2024cwr, Basilakos:2024diz}. Could this signal instead point to more exotic, cosmological sources? The challenge now lies in deciphering the nature of this background, determining whether it is primarily astrophysical or cosmological in origin, and what the nanohertz GW sky can ultimately reveal about the Universe's processes and gravity \cite{Chamberlin:2011ev, Powell:2019kid, Tasinato:2022xyq, Qin:2020hfy, Boitier:2021rmb, Chen:2021ncc, Chen:2021wdo, Wu:2021kmd, Liang:2021bct, Bernardo:2022vlj, Bernardo:2022rif, Zhu:2022bwf, Bernardo:2023mxc, Wu:2023pbt, Liang:2023ary, Liang:2023pbj, Cordes:2024oem, Liang:2024mex, Domenech:2024pow, Atkins:2024nvl}.

This moment represents an opportune time to revisit the foundational elements of both theory and PTA data analysis, ensuring that systematic errors do not obscure or exaggerate the detection of a SGWB signal \cite{DiMarco:2024irz, Agazie:2024stg, Goncharov:2024htb, Goncharov:2024fsi}. It is crucial to refine our theoretical frameworks and analytical techniques, which form the bedrock for interpreting PTA observations \cite{Bernardo:2024tde}. This review is dedicated to the theoretical aspects of this endeavor, providing a comprehensive exploration of the methods and principles necessary for a robust understanding of the SGWB as observed through PTAs (Sections \ref{sec:pulsar_timing_model}-\ref{sec:hellings_and_downs}). Our aim is to develop a theoretical playbook that addresses potential challenges in data interpretation, such as biases in the inferred spectrum and the impact of cosmic variance, while proposing new techniques for extracting maximum information from PTA datasets and broadening the field's scientific impact by suggesting new avenues for exploration (Section \ref{sec:beyond_hd_curve}). For readers interested in the broader astronomical context and the observational strategies that complement this theoretical foundation, we refer you to several recent reviews that thoroughly cover the observational and astrophysical aspects of PTA science \cite{Joshi:2013at, McLaughlin:2014wna, Manchester:2015mda, Lommen:2015gbz, Romano:2016dpx, Becker:2017yyc, Verbiest:2021kmt, Taylor:2021yjx, Verbiest:2024nid, Yunes:2024lzm}.

Before we proceed, we would like to acknowledge the extensive observational and theoretical efforts that have paved the way to where the field stands today. While the following is not an exhaustive list, we highlight several milestones that we believe have significantly advanced our theoretical understanding of a SGWB signal and its correlation in PTAs:
\begin{itemize}[label=\CheckmarkBold, leftmargin=*]
    \item 1978---Pulsar timing was proposed for the detection of nanohertz GWs \cite{Sazhin:1978myk, Detweiler:1979wn};
    \item 1983---The SGWB correlation was derived \cite{Hellings:1983fr}, and has since been known as the Hellings and Downs curve;
    \item 2001---The spectral profiles were drawn for SGWB sources 
    \cite{Phinney:2001di};
    \item 2002---Signal and source redshifts for low frequency GWs were analyzed \cite{Wyithe:2002ep};
    \item 2004---Signal from SMBHB was computed in a hierarchical cosmology \cite{Wyithe:2002ep, Sesana:2004sp};
    \item 2011---SGWB correlations were derived for non-Einsteinian GW polarizations propagating at the speed of light \cite{Chamberlin:2011ev};
    \item 2013---Correlations for the characterization of SGWB anisotropy were derived in real space \cite{Mingarelli:2013dsa};
    \item {2014---The angular power spectrum/harmonic space form of the HD correlation was introduced \cite{Gair:2014rwa};}
    \item {2016---Cosmic variance of the HD correlation was introduced in harmonic space redshift maps \cite{Roebber:2016jzl};}
    \item 2018---A power spectrum approach (PSA) was introduced for the calculation of SGWB correlations for non-Einsteinian GW modes \cite{Qin:2018yhy};
    \item 2019---Correlations for an anisotropic SGWB were derived in harmonic space \cite{Hotinli:2019tpc};
    \item 2020---Signatures of chirality in a SGWB signal were determined \cite{Belgacem:2020nda};
    \item 2020---PSA was generalized for subluminal SGWB correlations \cite{Qin:2020hfy};
    \item 2021---SGWB correlations in PTA were calculated for massive gravity \cite{Liang:2021bct};
    \item 2021---The PSA for luminal tensor GW modes was revisited, and generalized to finite pulsar distances \cite{Ng:2021waj};
    \item 2021---Correlations for an anisotropic polarized SGWB with finite pulsar distances were derived \cite{Liu:2022skj};
    \item 2022---The variance of HD was calculated \cite{Allen:2022dzg};
    \item 2022---HD variance was generalized for arbitrary pulsar distributions \cite{Allen:2022ksj};
    \item 2022---The PSA was generalized for subluminal GWs and non-Einsteinian modes in a SGWB \cite{Bernardo:2022rif};
    \item 2022---Variance of non-Einsteinian subluminal SGWB correlations were derived by the PSA \cite{Bernardo:2022xzl};
    \item {2023---Compelling evidence of the nanohertz SGWB by the PTAs \cite{NANOGrav:2023gor,Xu:2023wog,EPTA:2023fyk,Reardon:2023gzh};}
    \item 2023---Correlations for an anisotropic polarized SGWB for non-Einsteinian polarizations, subluminal GWs, and finite pulsar distances were obtained \cite{AnilKumar:2023yfw, Bernardo:2023jhs, AnilKumar:2023hza};
    \item 2024---Source statistics were characterized in a PTA observation \cite{Allen:2024rqk};
    \item {2024---Cumulants of the SGWB one- and two-point functions were introduced in PTA analysis \cite{Lamb:2024gbh, Bernardo:2024uiq};}
\end{itemize}
We must emphasize that there has been a surge of recent theoretical progress in the field that we highly recommend for reading \cite{Siemens:2006yp, Sesana:2008mz, Khmelnitsky:2013lxt, Porayko:2014rfa, Guzzetti:2016mkm, Caprini:2018mtu, Tahara:2020fmn, Adshead:2021hnm, Chu:2021krj, Garcia-Saenz:2022tzu, Tasinato:2022xyq, Bernardo:2022vlj, Acuna:2023bkm, Hwang:2023odi, Liang:2023ary, Caliskan:2023cqm, Domenech:2024pow, Depta:2024ykq, Inomata:2024kzr, Hu:2024wub}, and the exclusion of these works from our list absolutely does not diminish their impact.

The rest of this review proceeds as follows. In Section \ref{sec:pulsar_timing_model}, we briefly review the pulsar timing model and other relevant non-GW aspects, noises in a PTA observation to set the stage for the introduction to the SGWB signal in a PTA (Eq. \eqref{eq:timing_residual_correlation_general}). The two important components of the signal, the spectrum and the correlation, receive detailed attention next. In Section \ref{sec:smbhb_spectrum}, we derive the SGWB spectrum associated to coalescing circular SMBHBs (Eq. \eqref{eq:gw_density_smbhb}), revisiting Phinney's theorem \cite{Phinney:2001di}, and in Section \ref{sec:hellings_and_downs}, we derive the correlation due to a stochastic superposition of GWs, culminating to the HD curve (Eq. \eqref{eq:hd_curve}) \cite{Hellings:1983fr}. In Section \ref{sec:beyond_hd_curve}, we explore speculative and emerging theoretical aspects of the signal that have recently gained considerable progress, including the cosmic variance (Section \ref{subsec:cosmic_variance}), cumulants of the one- and two-point functions (Section \ref{subsec:cumulants}), subluminal GWs (Section \ref{subsec:subluminal_gws}), and anisotropy and polarization (Section \ref{subsec:anisotropy}).

\begin{table}[h!]
    \tbl{Common symbols used in this review.}
    {
    \scriptsize
    \renewcommand{\arraystretch}{1.4} 
    \centering
    \begin{tabular}{|c|c|}
    \hline
    SYMBOL & DESCRIPTION \\ \hline \hline
    $h_{ij}(t)$ & Gravitational Wave (GW) \\ \hline
    $r(t, \hat{e})$ & Pulsar Timing Residual \\ \hline
    $z(t, \hat{e})$ & Redshift Fluctuation \\ \hline
    $\varepsilon_{ij}$ & GW Polarization Basis Tensor \\ \hline
    $\hat{e}_a$ & Unit Vector---Earth to Pulsar $a$ \\ \hline
    $\hat{e}^i \otimes \hat{e}^j$ & Detector Tensor \\ \hline
    ${\cal P}(f)$ & Power Spectrum \\ \hline
    $\Omega_{\rm gw}(f)$ & GW Cosmological Density Parameter \\ \hline
    $\gamma(\hat{e}_a \cdot \hat{e}_b)$ & Overlap Reduction Function (ORF) \\ \hline
    $\zeta$ & Inter-pulsar Angular Separation \\ \hline
    $\gamma_{aa}$ & Auto-correlation/Auto-pulsar correlation \\ \hline
    $l$ & Multipole Number/Index \\ \hline
    $C_l$ & Angular Power Spectrum Multipole/Coefficient \\ \hline
    $D_a$ & Distance to Pulsar $a$ \\ \hline
    $f$ & GW Frequency \\ \hline
    $v$ & Group Velocity \\ \hline
    $\langle \cdots \rangle$ & Ensemble Average \\ \hline
    $\{ \cdots \}$ & Full-sky Average \\ \hline 
    $P_l(x)$ & Legendre Polynomial \\ \hline
    $j_l(x)$ & Spherical Bessel Function \\ \hline
    $Y_{lm}\left(\hat{k}\right)$ & Spherical Harmonics \\ \hline
    $\, _s Y_{lm}\left(\hat{k}\right)$ & Spin-weighted Spherical Harmonics \\ \hline
    $\vec{k} = k \hat{k}$ & Wave Vector \\ \hline
    $\tilde{h}_A(f, \hat{k})$ & GW Fourier Amplitude \\ \hline
    $D_{m'm}^l(-\alpha,-\theta,-\phi)$ & Wigner D Matrix \\ \hline
    $\Gamma(x)$ & Gamma Function \\ \hline
    $\, _2F_1(a, b;c;x)$ & Hypergeometric Function \\ \hline
    $\, _2\tilde{F}_1(a,b;c;x)$ & Regularized Hypergeometric Function $\, _2F_1(a, b;c;x)/\Gamma(c)$ \\ \hline
    $\, _2F_2(a, b;c, d;x)$ & Hypergeometric Function \\ \hline
    $\, _2\tilde{F}_2(a,b;c, d;x)$ & Regularized Hypergeometric Function $\, _2F_1(a, b;c, d;x)/\left( \Gamma(c) \Gamma(d) \right)$ \\ \hline
    \end{tabular}
    \label{tab:symbols}
    }
\end{table}

For a pedagogical introduction, we refer readers to Refs. \citen{Jenet:2014bea, Romano:2023zhb} and the NANOGrav tutorials in GitHub\footnote{\href{https://github.com/nanograv/15yr_stochastic_analysis}{https://github.com/nanograv/15yr\_stochastic\_analysis}}. We also encourage readers to explore free, open-source codes such as \texttt{TEMPO2} \cite{Hobbs:2006cd, Edwards:2006zg, Hobbs:2009yn}, \texttt{libstempo}, \texttt{ENTERPRISE} \cite{enterprise}, \texttt{PTAfast} \cite{2022ascl.soft11001B}, \texttt{PTArcade} \cite{Mitridate:2023oar}, and PTA data repositories\footnote{PTA data repositories 2024: \href{https://zenodo.org/communities/nanograv}{NANOGrav}, \href{https://zenodo.org/communities/epta}{EPTA/InPTA}, \href{https://data.csiro.au/collection/csiro:59374}{PPTA}, \href{https://gitlab.com/IPTA/DR2/tree/master/release}{IPTA}, \href{https://dmc.datacentral.org.au/dataset/meerkat-pulsar-timing-array-mpta-first-data-release}{Meerkat PTA}.} to corroborate the discussion in this review with simulations and to assist in research. It is worth emphasizing that we arbitrarily kept to the present NANOGrav constraints in Figures \ref{fig:red_noise}-\ref{fig:orf} for clarity of presentation and that public EPTA/InPTA, PPTA, and IPTA data sets were equally good options. Throughout, we use geometrized units ($G=c=1$) and the mostly-plus metric signature $(-,+,+,+)$. Table \ref{tab:symbols} lists the common symbols used in this text.

\section{The Pulsar Timing Model}
\label{sec:pulsar_timing_model}

We review the pulsar timing model (Section \ref{subsubsec:pulsar_timing_model}) and intrinsic stochastic noises (Section \ref{subsubsec:stochastic_part}) associated to a pulsar in a PTA \cite{1984JApA....5..369B, Hazboun:2019vhv, Antoniadis:2023lym, NANOGrav:2023hde}. Then, we proceed to the GW signal and correlation in the pulsar timing residuals (Sections \ref{subsec:gw_residual}-\ref{subsec:timing_and_orf}).

\subsection{Non-GW Pulsar Noises}
\label{subsec:non_gw_pulsar_noises}

\subsubsection{Pulsar timing model}
\label{subsubsec:pulsar_timing_model}

For a given pulsar $a$, the raw or pre-fit time-of-arrival (ToA) data, ${\mathbf t}_{a}$, is composed of two parts,
\begin{equation}
    {\mathbf t}_{a} = {\mathbf t}^{\rm det}_{a} + {\mathbf r}_a^{\rm stoc} \,,
\end{equation}
where `det' and `stoc' denote deterministic and stochastic (residual) pieces. The residual, $\delta{\mathbf t}_a$, is determined by marginalizing over the deterministic part using a timing model M, e.g., Eqs. (\ref{eq:r_TM}-\ref{eq:psi_TM}),
\begin{equation}
    {\mathbf t}_a^{\rm det}={\mathbf t}_a^{\rm M} + {\mathbf M}_a {\mathbf{\delta \xi}} \,,
\end{equation}
and subtracting the best fit out of the raw data,
\begin{equation}
    {\mathbf{\delta t}}_a = {\mathbf t}_a-{\mathbf t}_a^{\rm M} = {\mathbf r}_a^{\rm stoc} + {\mathbf M}_a \delta \xi \,,
\end{equation}
where $\xi$ are the timing model parameters and ${\mathbf M}_a=(\partial {\mathbf t}_a^{\rm det} / \partial {\mathbf \xi})_{ {\mathbf \xi}={\mathbf \xi}_0 }$ is a so-called design matrix. If the best fit timing model is a good representation of the deterministic part of the data, ${\mathbf \xi}_0\sim{\mathbf \xi}_{\rm true}$, then the residual will be left with only stochastic components, ${\mathbf{\delta t}}_a \sim {\mathbf r}_a^{\rm stoc}$. The resulting post-fit or timing model marginalized (TMM) residuals can be effectively described by a transmission function, ${\cal T}(f)$, encoding the overall sensitivity of a pulsar to stochastic signals \cite{Hazboun:2019vhv}. The quadratic spin-down model, accounting for a phase offset, the pulsar's spin period, and the period derivative, leads to a subnanohertz power suppression, reflected through the transmission function as ${\cal T}(f \lesssim 1 \ {\rm nHz}) \sim f^6$. We refer the readers to Ref. \citen{Hazboun:2019vhv} for other key features of the transmission function such as sky localization and parallax.

The subnanohertz power suppression in the TMM residuals motivate the use of other observables for detecting subnanohertz GWs. In Refs. \citen{DeRocco:2022irl, DeRocco:2023qae}, it was shown that the pulsar parameter drifts turn out to be a suitable observable for subnanohertz GW detection. Astrometric observations are also a promising candidate for constraining subnanohertz GWs \cite{Book:2010pf, Klioner:2017asb, Darling:2018hmc, Jaraba:2023djs, Caliskan:2023cqm}.

Uncertainties of the parameters in a traditional timing model used to describe a pulsar's ToA data
lead to a timing residual given by \cite{1984JApA....5..369B}
\begin{equation}
\label{eq:r_TM}
{\mathbf {\delta t}}_a^{\rm det}={\mathbf M}_a {\mathbf{\delta \xi}}=\sum_{i=0}^7 {\mathbf {\delta \xi}}_i \psi_i(t)\,,
\end{equation}
where the $\psi_i(t)$'s are
\begin{align}
&\psi_0(t)=1,\quad \psi_1(t)=t,\quad \psi_2(t)=t^2, \nonumber \\
&\psi_3(t)=\sin({\bar\omega}t),\quad
\psi_4(t)=\cos({\bar\omega}t),\quad
\psi_5(t)=t\cos({\bar\omega}t),
\nonumber \\
&\psi_6(t)=t\sin({\bar\omega}t),\quad
\psi_7(t)=\sin(2{\bar\omega}t)\,, \label{eq:psi_TM}
\end{align}
and ${\bar\omega}=2\pi/{\rm yr}$.
$\psi_{i=0,1,2}(t)$ represent a quadratic spin-down model, fitting to the phase offset, spin period, and 
period derivative, respectively. 
This model can be compared to a kinematic equation, $s=s_0+v_0 t+a t^2/2$,
where $s_0$, $v_0$, and $a$ are initial distance, initial speed, and constant acceleration, respectively, with
$v_0$ being analogous to the redshift $z(t)$ and $s$ to the timing residual 
$r(t)=\int dt' z(t')$ in Eq.~(\ref{eq:timing_residual}). 
$\psi_{i=3,4,5,6}(t)$ are fitted to the sky position of the pulsar with a frequency of 1/yr, 
corresponding to the Earth’s yearly orbital motion around the Sun. 
$\psi_7(t)$ with a frequency of 2/yr corresponds to a parallax measurement of the pulsar distance. We refer the readers to Refs. \citen{1984JApA....5..369B, Manchester:2015mda, Taylor:2021yjx} for more details about the timing model and to \texttt{TEMPO2} \cite{Hobbs:2006cd, Edwards:2006zg} for an engaging software.

\subsubsection{The stochastic part of the residual}
\label{subsubsec:stochastic_part}

Like any time series, the pulsar timing residual can be expanded as a Fourier series \cite{Taylor:2021yjx},
\begin{equation}
\label{eq:residual_fourier_series}
r(t,\hat{e})= {\alpha_0(\hat{e})\over2}+\sum_{k=1}^{\infty} \alpha_k(\hat{e}) \sin(\omega_k t) + \sum_{k=1}^{\infty} \beta_k(\hat{e}) \cos(\omega_k t)\,,
\end{equation}
where $\omega_k = 2 \pi f_k$, $f_k=k/T$, $k$ is a positive integer, $T$ is the total observation time, and ${\hat e}$ is a pulsar direction unit vector. The Fourier components of the timing residual are given by
\begin{eqnarray}
\alpha_0(\hat{e}) &=& \dfrac{2}{T} \int_0^T dt\, r(t,\hat{e})\,, \nonumber \\
\alpha_i(\hat{e}) &=& \dfrac{2}{T}\int_0^T dt\, r(t,\hat{e}) \sin(\omega_i t) \,,\nonumber \\
\beta_i(\hat{e}) &=& \dfrac{2}{T}\int_0^T dt\, r(t,\hat{e}) \cos(\omega_i t) \,.
\end{eqnarray}
By referring to the Fourier bins/components, $\alpha(\hat{e})$ and $\beta(\hat{e})$, we can distinguish clearly pulsar intrinsic noises with a GW signal.

The pulsar intrinsic noises can be described briefly. For a white measurement noise, the variances of the Fourier components of the PTA residuals are characterized by
\begin{equation}
\langle \alpha_i(\hat{e}_a) \alpha_j (\hat{e}_b)\rangle =
\langle \beta_i(\hat{e}_a) \beta_j(\hat{e}_b) \rangle \sim \sigma_a^2\,\delta_{ab}\,\delta_{ij}\,,
\end{equation}
where $\sigma_a$ is a constant width specific to each pulsar. For the intrinsic red noise \cite{2023arXiv230300931R}, the variances of the Fourier components of the PTA residuals are given by
\begin{equation}
\label{eq:pulsar_red_noise}
\langle \alpha_i(\hat{e}_a) \alpha_j (\hat{e}_b)\rangle =
\langle \beta_i(\hat{e}_a) \beta_j(\hat{e}_b) \rangle \sim 
\frac{2A_a^2}{\pi^2} \frac{1}{Tf_{\rm ref}^3} \left(\frac{f_i}{f_{\rm ref}}\right)^{-\gamma_a}\,\delta_{ab}\,\delta_{ij},
\end{equation}
where the power spectrum (unique to each pulsar) is assumed to be a power law with some spectral index $\gamma$, $f_{\rm ref}\sim {\cal O}(1) \ {\rm yr}^{-1}$ is a reference frequency, and $T$ is the duration of the observation \cite{vanHaasteren:2012hj, vanHaasteren:2014qva}.

\begin{figure}[h!]
    \centering
    \includegraphics[width=1.0\textwidth]{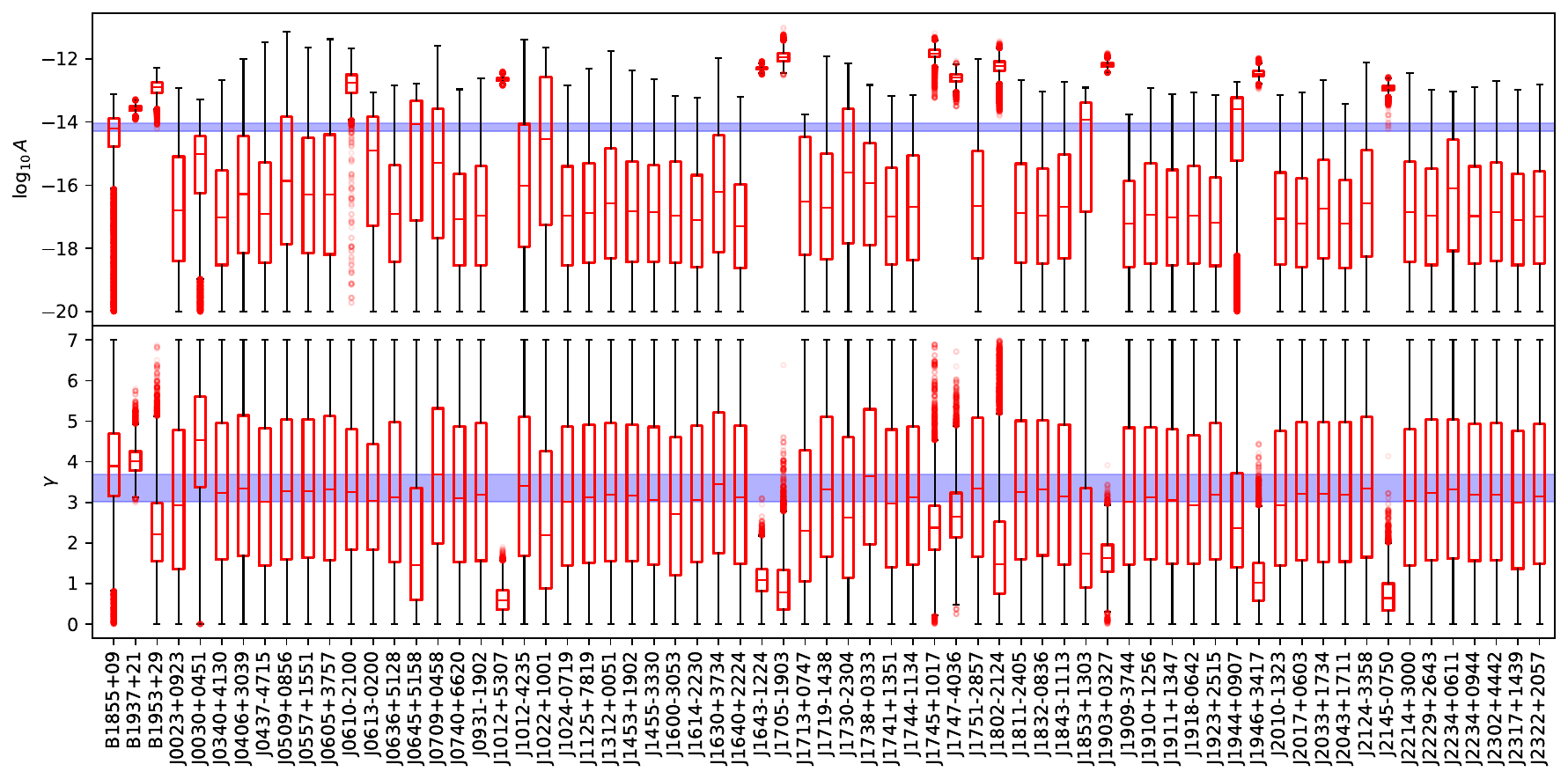}
    \caption{Red noise parameters in the NANOGrav 15-year data set (red whiskers) and the corresponding SGWB power spectrum (blue horizontal bands) \cite{NANOGrav:2023gor}.}
    \label{fig:red_noise}
\end{figure}

It is worth mentioning that the pulsar intrinsic red noise time scales are similar to nanohertz GWs. In PTA analysis, both the SGWB and pulsar red noise are associated with a red (time-dependent) process and, hence, a power spectral density via Wienner-Khinchin theorem \cite{vanHaasteren:2012hj}. In this case, the SGWB is perceived as a common spatially-correlated process across pulsars in contrast with the uncorrelated intrinsic red noise. Figure \ref{fig:red_noise} shows the typical intrinsic red noise parameters and that of a SGWB in a PTA. There are also features of the observation that can factor in as a systematic correlation between pulses, such as clock errors and ephemeris issues. For a more in-depth discussion, we refer the readers to Refs. \citen{2016MNRAS.455.4339T, Roebber:2019gha, NANOGrav:2020tig, Taylor:2021yjx}.

\subsection{Pulsar Timing Residual from GWs}
\label{subsec:gw_residual}

In this section, we focus on the GW part of the pulsar timing residual.

We start with a stochastic superposition of GWs, with various frequencies $f$ and propagation direction $\hat{k}$,
\begin{equation}
\label{eq:gw_general}
    h_{ij}\left(\eta, \vec{x}\right) = \sum_{A={+,\times}} \int_{-\infty}^\infty df \int_{S^2} d\hat{k} \ h_A\left(f, \hat{k}\right) \varepsilon_{ij}^A e^{-2\pi i f \left( \eta - \hat{k} \cdot \vec{x} \right)} \,,
\end{equation}
where $\eta$ is a conformal time, $\vec{x}$ is a position vector, and $\varepsilon_{ij}^+$ and $\varepsilon_{ij}^\times$ are the transverse-traceless polarization basis tensors. For a GW propagating along the $\hat{z}$ direction, the basis can be written in Cartesian coordinates $\left( \hat{x}, \hat{y}, \hat{z} \right)$ as
\begin{equation}
    \varepsilon^{+} = 
    \left(
    \begin{array}{ccc}
    1 & 0 & 0 \\
    0 & -1 & 0 \\
    0 & 0 & 0 
    \end{array}
    \right) \ \ \ \ 
    {\rm and} \ \ \ \ \varepsilon^{\times} = 
    \left(
    \begin{array}{ccc}
    0 & 1 & 0 \\
    1 & 0 & 0 \\
    0 & 0 & 0 
    \end{array}
    \right) \,.
\end{equation}
In general, for a GW propagating along the $\hat{\Omega}$ direction, the polarization basis tensors can be expressed as $\varepsilon^{+} = \hat{m} \otimes \hat{m} - \hat{n} \otimes \hat{n}$ and $\varepsilon^{\times} = \hat{m} \otimes \hat{n} + \hat{n} \otimes \hat{m}$ 
where $\left( \hat{m}, \hat{n}, \hat{\Omega} \right)$ form an orthonormal basis \cite{Boitier:2021rmb}. Now, the effect of a GW on the pulsar timing residual, $r(t, \hat{e})$, is canonically quantified by computing the GW contribution at the location of the Earth and the pulsar, and subtracting the results. Alternatively, this can be done by integrating over the redshift space fluctuations, $z(t, \hat{e})$,
\begin{equation}
\label{eq:timing_residual}
    r\left(t, \hat{e}\right) = \int_0^t dt' \ z\left(t', \hat{e}\right) \,,
\end{equation}
where $t$ is the time of an observation. For a GW, e.g., $h_{ij}\left(\eta, \vec{x}\right)$ given by Eq. \eqref{eq:gw_general}, and photons emitted at a time $\eta_e$ and received by the detector at time $\eta_r$, the redshift space fluctuation can be written as the Sachs-Wolfe integral \cite{Sachs:1967er},
\begin{equation}
\label{eq:z_swolf}
    z\left(t', \hat{e}\right) = - \dfrac{1}{2} \int_{t' + \eta_e}^{t' + \eta_r} d\eta \ \left( \hat{e}^i \otimes \hat{e}^j \right) \ \partial_\eta h_{ij} \left( \eta, \vec{x} \right) \,.
\end{equation}
The quantity $\hat{e}^i \otimes \hat{e}^j$ is often referred to as `detector tensor' $d_{ij}$.

\begin{figure}[h!]
    \centering
    \includegraphics[width=0.975\textwidth]{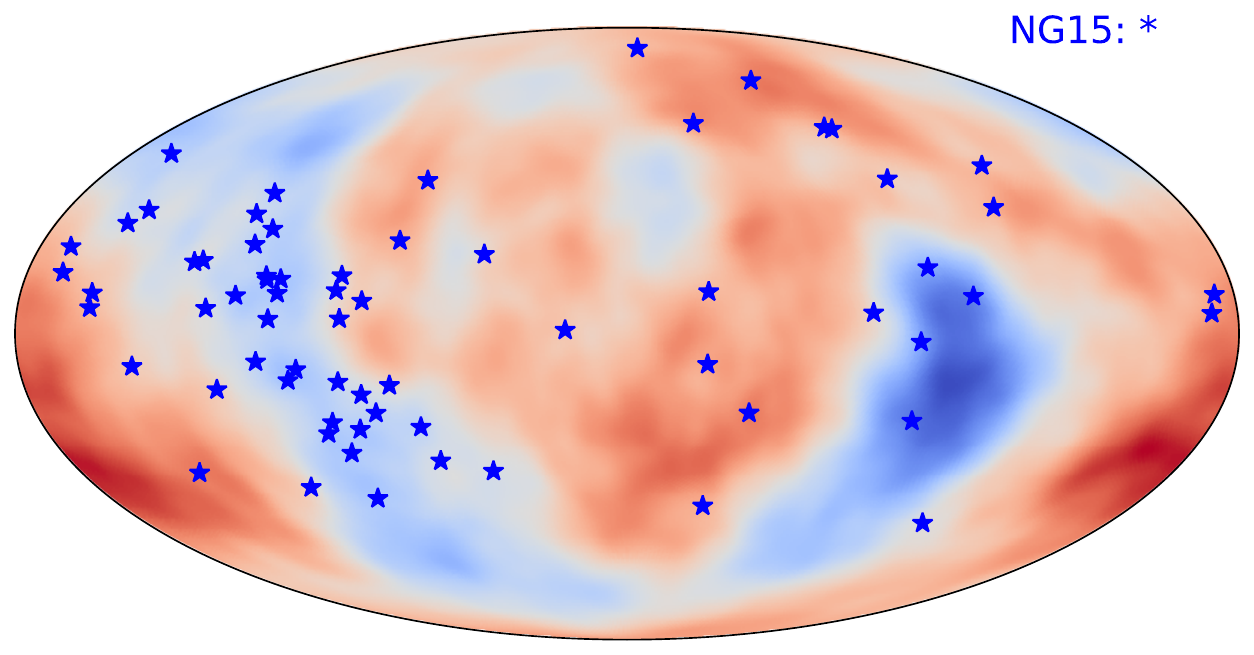}
    \caption{Sky distribution of pulsars in the NANOGrav 15-year data set on a projection of a SGWB \cite{NANOGrav:2023hde}.}
    \label{fig:pulsars}
\end{figure}

Substituting Eq. \eqref{eq:gw_general} into Eqs. (\ref{eq:timing_residual}-\ref{eq:z_swolf}), we obtain
\begin{equation}
\begin{split}
    r\left(t, \hat{e}\right) = & \int_0^t dt' \int_{t' + \eta_e}^{t' + \eta_r} d \eta  \sum_{A} \int_{-\infty}^\infty df \int_{S^2} d\hat{k} \\
    & \ \ \ \ \times \pi i f \left( \hat{e}^i \otimes \hat{e}^j \right) h_A \left(f, \hat{k}\right) \varepsilon_{ij}^A\left(\hat{k}\right) e^{-2\pi i f \left( \eta - \hat{k} \cdot \vec{x} \right)} \,.
\end{split}
\end{equation}
By expanding the plane wave in spherical harmonics, $Y_{lm}\left(\hat{e}\right)$,
\begin{equation}
    e^{2\pi i f \hat{k}\cdot\vec{x}} = 4\pi \sum_{lm} i^l j_l\left(2\pi f|\vec{x}|\right) Y_{lm}^*\left(\hat{k}\right) Y_{lm}\left(\hat{e}\right) \,,
\end{equation}
then we are able to recast the last result into
\begin{equation}
\begin{split}
    r\left(t, \hat{e}\right) = & \int_0^t dt' \int_{t' + \eta_e}^{t' + \eta_r} d \eta  \sum_{A} \int_{-\infty}^\infty df \int_{S^2} d\hat{k} \ 4 \pi^2 i f  \left( \hat{e}^i \otimes \hat{e}^j \right) h_A \left(f, \hat{k}\right) \varepsilon_{ij}^A\left(\hat{k}\right) \\
    & \ \ \ \ \ \ \ \ \times  e^{-2\pi i f \eta} \sum_{lm} i^l j_l\left(2\pi f \left( t' + \eta_r - \eta \right) \right) Y_{lm}^*\left(\hat{k}\right) Y_{lm}\left(\hat{e}\right) \,,
\end{split}
\end{equation}
where $\vec{x} = \left( t' + \eta_r - \eta \right) \hat{e}$ is the position vector to the pulsar at time $t'$. The $j_l(x)$'s are the spherical Bessel functions of the first kind. To simplify the result further, we consider the following identities,
\begin{equation}
    \int_0^t dt' \int_{t' + \eta_e}^{t' + \eta_r} d\eta' e^{-i \omega \eta'} W \left( t' + \eta_r - \eta' \right) = \dfrac{1 - e^{-i \omega t}}{i \omega} \int_{\eta_e}^{\eta_r} d\eta \ e^{-i \omega \eta} W\left( \eta_r - \eta \right)
\end{equation}
and
\begin{equation}
    \int_{\eta_e}^{\eta_r} d\eta e^{- i \omega \eta} j_l \left( \omega \left(\eta_r - \eta\right) \right) = \dfrac{e^{-i \omega \eta_r}}{\omega} \int_0^{\omega D} dx \ e^{i x} j_l\left(x\right) \,,
\end{equation}
where $W(t)$ is an arbitrary function. By using the above expression, then the pulsar timing residual can now be written as a Laplace/harmonic series,
\begin{equation}
\label{eq:gwb_residual_harmonic_series}
    r\left(t, \hat{e}\right) = \sum_{lm} a_{lm} (t) Y_{lm} \left( \hat{e} \right) \,,
\end{equation}
where the coefficients $a_{lm}(t)$'s are given by
\begin{equation}
\label{eq:gwb_residual_harmonic_coefficients}
\begin{split}
    a_{lm}(t) = & \sum_{A=+,\times} \int_{-\infty}^\infty d\omega \int_{S^2} d\hat{k} \ \left( 1 - e^{-i \omega t} \right) \left( \dfrac{e^{- i \omega \eta_r}}{\omega} \right) \\
    & \ \ \ \ \times \left( \hat{e}^i \otimes \hat{e}^j \right) h_A\left(\dfrac{\omega}{2\pi}, \hat{k}\right) \varepsilon_{ij}^A\left(\hat{k}\right) \int_0^{\omega D} dx \ e^{ix} i^l j_l\left(x\right) Y^*_{lm}\left(\hat{k}\right) \,.
\end{split}
\end{equation}
Notice that in the last steps, we wrote $\omega=2\pi f$. The result (Eqs. (\ref{eq:gwb_residual_harmonic_series}-\ref{eq:gwb_residual_harmonic_coefficients})) is best visualized in spherical coordinates, as shown in Figure \ref{fig:pulsars}, with the location and timing data for each pulsar given by $\hat{e}$ and $a_{lm}(t)$, respectively.

In the next section, we use the last result to compute the pulsar timing residual correlation.

\subsection{The Timing Residual Correlation due to a SGWB}
\label{subsec:timing_and_orf}

Consider a Gaussian and isotropic SGWB,
\begin{equation}
\label{eq:gwb_power_spectrum}
    \langle h_A\left(f, \hat{k}\right) h_{A'}^* \left( f', \hat{k}'\right) \rangle = {\cal P}(f) \delta_{AA'} \delta \left( f - f' \right) \delta \left( \hat{k} - \hat{k}' \right) \,,
\end{equation}
where $\langle \cdots \rangle$ denotes an ensemble average, and ${\cal P}(f)$ is the SGWB power spectrum.

Using the Eqs. (\ref{eq:gwb_residual_harmonic_series}-\ref{eq:gwb_residual_harmonic_coefficients}), the timing residual cross correlation between a pair of pulsars can be written as
\begin{equation}
    \langle r\left(t_a, \hat{e}_a\right) r\left(t_b, \hat{e}_b\right) \rangle = \sum_{l_1 m_1} \sum_{l_2 m_2} \langle a_{l_1 m_1}(t_a) a^*_{l_2 m_2}(t_b) \rangle Y_{l_1 m_1}\left( \hat{e}_a \right) Y^*_{l_2 m_2}\left( \hat{e}_b \right) \,,
\end{equation}
where
\begin{equation}
\label{eq:two_point_lm}
\begin{split}
    \langle a_{l_1 m_1}(t_a) a^*_{l_2 m_2}(t_b) \rangle = & \int_{-\infty}^\infty \dfrac{df}{\left(2\pi f\right)^2} C(f, t_a, t_b) \sum_{A_1 A_2} \int_{S^2} d\hat{k} \\
    & \ \ \ \ \times {\cal P}\left( f \right) J^{A_1}_{l_1 m_1} \left( f D_a, \hat{k} \right) J^{A_2 *}_{l_2 m_2} \left( f D_b, \hat{k} \right) \,,
\end{split}
\end{equation}
\begin{equation}
\label{eq:C_filter_def}
C(f,t_a, t_b)=\left( 1 - e^{-2\pi i f t_a} \right) \left( 1 - e^{2\pi i f t_b} \right) \,,
\end{equation}
and
\begin{equation}
\label{eq:Jlm_def}
\begin{split}
    J_{lm}^A \left( fD, \hat{k} \right) = & 2\pi \int_0^{2\pi f D} dx \ e^{i x} \sum_{LM} i^L Y^*_{LM} \left( \hat{k} \right) j_L(x) \\
    & \ \ \ \ \times \int_{S^2} d\hat{e} \ \left( \hat{e}^i \otimes \hat{e}^j \right) \varepsilon_{ij}^A\left(\hat{k}\right) Y_{LM}\left( \hat{e} \right) Y_{lm}^*\left(\hat{e}\right) \,.
\end{split}
\end{equation}
After some manipulation, the correlation can be written as
\begin{equation}
\label{eq:timing_residual_correlation_general}
    \langle r\left(t_a, \hat{e}_a\right) r\left(t_b, \hat{e}_b\right) \rangle = \int_{-\infty}^\infty \dfrac{df}{\left(2\pi f\right)^2} C(f, t_a, t_b) {\cal P}(f) \sum_{A=+,\times}\gamma^A(fD_a, fD_b, \hat{e}_a\cdot\hat{e}_b) \,,
\end{equation}
where the $\gamma^A(fD_a, fD_b, \hat{e}_a\cdot\hat{e}_b)$'s are given by
\begin{equation}
\label{eq:orf_general}
\begin{split}
    \gamma^A \left(f D_a, fD_b, \hat{e}_a\cdot\hat{e}_b \right) = & \sum_{l_1 m_1} \sum_{l_2 m_2} Y_{l_1 m_1}\left( \hat{e}_a \right) Y^*_{l_2 m_2} \left( \hat{e}_b \right) \\
    & \ \ \times \int_{S^2} d\hat{k} \ Y_{00}\left(\hat{k}\right) J^A_{l_1 m_1} \left( f D_a, \hat{k} \right) J^{A*}_{l_2 m_2} \left( f D_b, \hat{k} \right) \,.
\end{split}
\end{equation}
Note that we have utilized isotropy to assume that $\gamma^A \left(f D_a, fD_b, \hat{e}_a, \hat{e}_b \right) = \gamma^A \left(f D_a, fD_b, \hat{e}_a\cdot\hat{e}_b \right)$. In simple words, for an isotropic SGWB, the timing residual cross correlation between a pair of Galactic millisecond pulsars depend only on the angular separation. The spatial correlation of the SGWB signal given by Eq. \eqref{eq:orf_general} is visualized in Figure \ref{fig:pulsars} in the background of pulsars.

The result of this section (Eq. \eqref{eq:timing_residual_correlation_general}) shows that the SGWB part of the timing residual correlation is described by three quantities, a time-frequency filter $C(f,t,t')$, a power spectrum, ${\cal P}(f)$, and a spatial correlation, $\gamma \left(f D_a, fD_b, \hat{e}_a\cdot\hat{e}_b \right)$. The factor $C(f,t,t')$ is associated with the GW temporal evolution and has important implications for correctly interpreting a SGWB signal in PTAs \cite{Bernardo:2024tde}. The latter two will be derived and described in detail in Sections \ref{sec:smbhb_spectrum}-\ref{sec:hellings_and_downs}.

When interpreted through Eq. \eqref{eq:residual_fourier_series}, the SGWB signal, Eq. \eqref{eq:timing_residual_correlation_general}, can be recognized as a convolution between the power spectrum, the spatial correlation, and filters, specific to every frequency bin and component of the correlation \cite{Bernardo:2024tde}. In the following, we derive the filters for the $\langle \alpha_i(\hat{e}_a) \alpha_{j}(\hat{e}_b) \rangle$ and $\langle \beta_i(\hat{e}_a) \beta_{j}(\hat{e}_b) \rangle$. Using Eqs. \eqref{eq:timing_residual_correlation_general}, we obtain
\begin{align}
\langle \alpha_i(\hat{e}_a) \alpha_j(\hat{e}_b) \rangle&= \left(2\over T\right)^2 \int_0^T \int_0^Tdt dt' 
\langle r(t,\hat{e}_a)r(t',\hat{e}_b) \rangle \sin(\omega_i t) \sin(\omega_j t') \nonumber \\
&=  \int_{f_{\rm min}}^{f_{\rm max}} \frac{2\,df}{(2\pi f)^2}\, {\cal C}_{ij}^\alpha (f) I(f) \gamma(fD,\hat{e}_a,\hat{e}_b)\,,
\label{aiaj}
\end{align}
\begin{align}
\langle \beta_i(\hat{e}_a) \beta_j(\hat{e}_b) \rangle&= \left(2\over T\right)^2 \int_0^T \int_0^Tdt dt'
\langle r(t,\hat{e}_a)r(t',\hat{e}_b) \rangle \cos(\omega_i t) \cos(\omega_j t') \nonumber \\
&=  \int_{f_{\rm min}}^{f_{\rm max}} \frac{2\,df}{(2\pi f)^2}\, {\cal C}_{ij}^\beta (f) I(f) \gamma(fD,\hat{e}_a,\hat{e}_b)\,,
\label{bibj}
\end{align}
where we have defined the frequency filters as
\begin{align}
{\cal C}_{ij}^\alpha (f)&=\left(2\over T\right)^2 \int_0^T \int_0^Tdt dt' \,C(f,t,t') \sin(\omega_i t) \sin(\omega_j t')\,, \\
{\cal C}_{ij}^\beta (f)&=\left(2\over T\right)^2 \int_0^T \int_0^Tdt dt' \,C(f,t,t') \cos(\omega_i t) \cos(\omega_j t')\,.
\end{align}
Note that we have truncated the integral to $f_{\rm min}\sim1/T$ to $f_{\rm max}={\cal O}(10)/T$, corresponding to the PTA sensitivity range. Closing the integral (substituting Eq. \eqref{eq:C_filter_def}), we obtain the filters
\begin{align}
\label{eq:alpha_filter}
{\cal C}_{ij}^\alpha (f) &= \frac{4 i j \sin ^2(\pi  f T)}{\pi ^2 \left(i^2-f^2 T^2\right) \left(j^2-f^2 T^2\right)}\,, \\
\label{eq:beta_filter}
{\cal C}_{ij}^\beta (f) &= \frac{4 f^2 T^2 \sin ^2(\pi  f T)}{\pi ^2 \left(i^2-f^2 T^2\right) \left(j^2-f^2 T^2\right)}\,,
\end{align}
which we refer to as low-pass and high-pass filters, respectively \cite{Bernardo:2024tde}. By tracing the above steps, it is easy to show that the filter for $\langle \alpha_i(\hat{e}_a) \beta_{j}(\hat{e}_b) \rangle$ vanishes, meaning that the $\alpha$- and $\beta$-Fourier bins (Eq. \eqref{eq:residual_fourier_series}) are uncorrelated. It is worth noting that the filters imply that different frequency bins, $i \neq j$, are very weakly correlated, or not completely independent. Above all, the filters (Eqs. (\ref{eq:alpha_filter}-\ref{eq:beta_filter})) show that the $\alpha$- and $\beta$-bins have different variances/spectra, which if not properly accounted, could lead to systematic errors on the inferred SGWB parameters in a PTA \cite{Bernardo:2024tde}.

\section{The Power Spectrum of Supermassive Black Hole Binaries}
\label{sec:smbhb_spectrum}

In this section, we revisit the standard circular SMBHB spectrum considered in the standard interpretation of a GW signal in PTA.

We start with Phinney's theorem \cite{Phinney:2001di}. By definition, the total energy density in the form of GWs, ${\cal E}_{\rm {gw}}$, is given by
\begin{equation}
\label{eq:Egw_def}
{\cal E}_{\rm {gw}} = \int d (\log f) \Omega_{\rm gw}(f) \left( \rho_{\rm c} c^2 \right) \,,
\end{equation}
where $f$ are observed frequencies, $\rho_{\rm c}$ is the critical energy density of the Universe, and $\Omega_{\rm gw}(f)$ denotes the fraction of energy in GWs relative to $\rho_{\rm c}$ in a logarithmic frequency bin. To relate the density parameter $\Omega_{\rm gw}(f)$ to binaries, and subsequently the power spectrum ${\cal P}(f)$, we consider an innumerable number of radiating sources, emitting GWs with energies $df_r(dE/df_r)$ in the frequencies between $f$ and $f + df$ and a comoving number density $N(z)$ where $z$ is the cosmological redshift. In the source rest frame, the emitted frequencies are $f_{r}=f(1+z)$, i.e., radiation is cosmologically-redshifted before reception at observatories. Then, the total energy density, counting the radiating sources and invoking isotropy and homogeneity, can be written as
\begin{equation}
{\cal E}_{\rm {gw}} = \int dz N(z) \int d f_r \dfrac{dE}{df_r} \left( \dfrac{1}{1+z} \right) \,.
\end{equation}
The last factor, $1/(1+z)$,  accounts for the redshifting of the sources' radiation. The above expression can be easily recast to
\begin{equation}
\label{eq:Egw_sources}
{\cal E}_{\rm {gw}} = \int dz N(z) \int d (\log f) \dfrac{dE}{d (\log f_r)} \left( \dfrac{1}{1+z} \right) \,.
\end{equation}
By comparing \eqref{eq:Egw_def} and \eqref{eq:Egw_sources} we obtain an expression for the GW density parameter
\begin{equation}
\label{eq:Omega_gw_theorem}
\Omega_{\rm gw}(f) \left( \rho_{\rm c} c^2 \right) = \int dz N(z) \dfrac{dE}{d (\log f_r)} \left( \dfrac{1}{1+z} \right) \,.
\end{equation}
This is Phinney's theorem \cite{Phinney:2001di}. For generalizations, for modified dispersion relations and one-point cumulants, please see Refs. \citen{Lamb:2024gbh, Cruz:2024esk}.

For circular SMBHBs, the leading (Newtonian) term \cite{Mingarelli:2012hh, Susobhanan:2020iyj, Susobhanan:2022nzv, DeFalco:2023djo, DeFalco:2024ojf} in the GW spectrum of a single source is $dE/d (\log f_r)=(\pi G f_r)^{2/3} {\cal M}^{5/3}/3$, where ${\cal M}$ is the chirp mass, ${\cal M}^{5/3}=M_1 M_2 (M_1+M_2)^{-1/3}$, and $M_1$, $M_2$ are the masses of the SMBHs. Note that this requires no knowledge of general relativity (GR) except that the GW frequency is twice the orbital frequency of the binary. Substituting this into Eq. \eqref{eq:Omega_gw_theorem}, we obtain
\begin{equation}
\Omega_{\rm gw}(f) \left( \rho_{\rm c} c^2 \right) = \dfrac{(\pi G)^{2/3}}{3} f^{2/3} {\cal M}^{5/3} \int dz \dfrac{ N(z) }{ (1+z)^{1/3} } \,.
\end{equation}
For an equal mass binary, $M_1=M_2=M$, ${\cal M}=3^{3/5} M \sim 0.87 M$, and that the critical energy density of the Universe is given by $\rho_{\rm c} c^2 =3H_0^2/(8\pi G)\sim 2.78 \times 10^{11} h^2 M_{\odot} c^2 \ {\rm Mpc}^{-3}$ where $h=H_0/(100 \ {\rm km} \ {\rm s}^{-1}{\rm Mpc}^{-1})$. It is useful to write the gravitational constant as $G\sim 4.30\times10^{-9}$ km$^2$ Mpc $M_{\odot}^{-1}$s$^{-2}\sim 4.52 \times 10^{-48}$ Mpc$^3$ $M_{\odot}^{-1}$s$^{-2}$. Putting everything together, with $M \sim 10^7 M_{\odot}$, we obtain
\begin{equation}
\label{eq:gw_density_smbhb}
\Omega_{\rm gw}(f) h^2 \sim \left(3.47 \times 10^{-9}\right) \left( \dfrac{f}{1 \ {\rm yr}^{-1}} \right)^{2/3}  \left( \dfrac{ {\cal M} }{ 10^{7} M_{\odot} } \right)^{5/3} \int \dfrac{dz}{ (1 + z)^{1/3} } \left( \dfrac{N(z)}{ 1 \ {\rm Mpc}^{-3} } \right) \,.
\end{equation}
For typical values, $f\sim 1$ yr$^{-1}$, ${\cal M}\sim 10^7 M_{\odot}$, and $N(z)\sim N_0\sim 10^{-4}$ Mpc$^{-3}$, this gives the rough estimate $\Omega_{\rm gw}(f) h^2 \sim 10^{-13}$.

We next relate the GW cosmological energy density parameter, $\Omega_{\rm gw}(f)$, to the SGWB power spectrum, ${\cal P}(f)$. This can be teased out with Eq. \eqref{eq:gwb_power_spectrum}, leading to the relation \cite{Taylor:2021yjx}
\begin{equation}
\label{eq:psd_to_density}
\Omega_{\rm gw} (f) = \dfrac{ d (\rho_{\rm gw}/\rho_{\rm c}) }{ d (\log f) } = \dfrac{2\pi^2}{3 H_0^2} f^3 S(f) \,,
\end{equation}
where the one-sided power spectral density $S(f)$ is related to the power spectrum as ${\cal P}(f)=S(f) / (16 \pi)$.

Another related quantity is the GW characteristic strain, defined as $h_{\rm gw}(f)=\sqrt{f S(f)}$. The strain $h_{\rm gw}$ is typically parametrized as
\begin{align}\label{AIdef}
    h_{\rm gw}(f) = A_{\rm gw}\left(\frac{f}{f_{\rm ref}}\right)^\alpha\,.
\end{align}
GW strains relevant in PTAs have $h_{\rm gw}\sim A_{\rm gw}\sim 10^{-15}$. Comparing this with Eq. \eqref{eq:psd_to_density}, we obtain $\alpha=-2/3$ for circular SMBHBs. PTA observations directly constrain $\gamma=3-2\alpha$. This gives the standard spectral index $\gamma=13/3$ for circular SMBHBs. For a scale invariant power spectrum, $\Omega_{\rm gw}(f)\sim$ constant, due to primordial tensor fluctuations, this gives $\alpha=-1$ and $\gamma=5$ \cite{Taylor:2021yjx}. For a general power law strain/power spectrum, this gives $\Omega_{\rm gw}(f)\sim f^{2\alpha + 2}$ and $S(f)\sim {\cal P}(f)\sim f^{2\alpha-1}=f^{2-\gamma}$.

\begin{figure}[h!]
    \centering
    \includegraphics[width=0.975\textwidth]{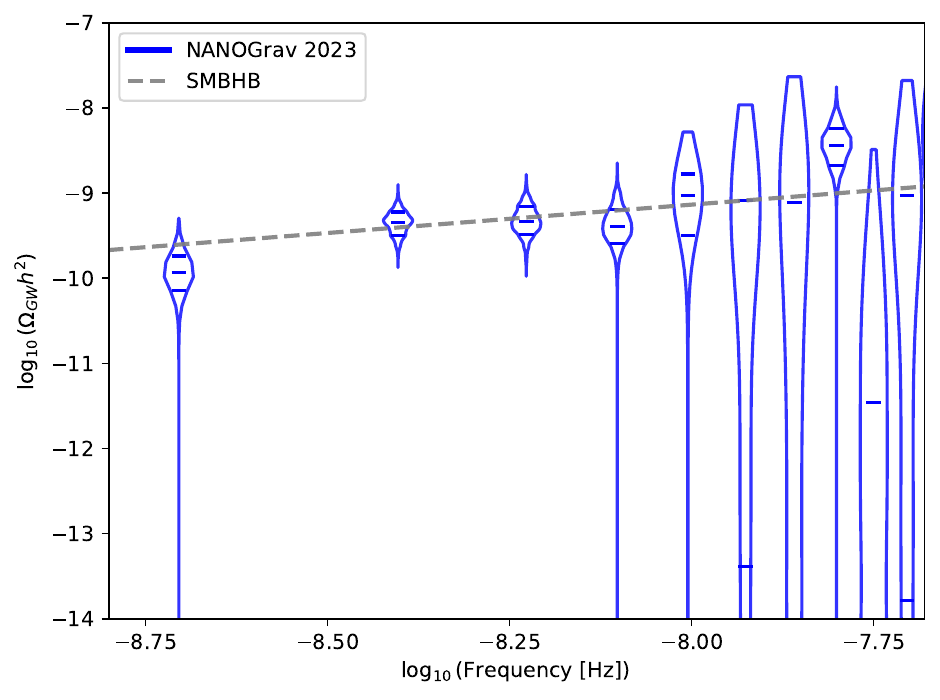}
    \caption{An inferred common spectrum in the NANOGrav 15-year data set \cite{NANOGrav:2023gor} (blue violins) and the SMBHB spectrum (gray dashed line).}
    \label{fig:spectrum}
\end{figure}

Figure \ref{fig:spectrum} shows a common spectrum process across pulsars attributed to a SGWB, together with the SMBHB spectrum. The widths of the uncertainty motivate a couple of interesting cosmological sources of the observed signal \cite{EPTA:2023xxk, NANOGrav:2023hvm, Wu:2023hsa, Ellis:2023oxs, Bian:2023dnv, Bi:2023tib, Verbiest:2024nid}. Post-Newtonian and nonlinear corrections can also be expected to be constrained in the signal when individual binaries become resolvable \cite{Mingarelli:2012hh, Susobhanan:2020iyj, Susobhanan:2022nzv, DeFalco:2023djo, DeFalco:2024ojf}.

\section{The Hellings and Downs Curve}
\label{sec:hellings_and_downs}

This section deals with the spatial correlation, $\gamma \left(f D_a, fD_b, \hat{e}_a\cdot\hat{e}_b \right)$, of the SGWB signal in a PTA. In Section \ref{subsec:Jlm_tensor}, we derive in detail the $J_{lm}^A(fD, \hat{k})$'s for GWs (Eq. \eqref{eq:Jlm_def}), and in Section \ref{subsec:power_spectra_tensor}, we obtain the so-called Hellings and Downs curve \cite{Hellings:1983fr, Romano:2016dpx}.

\subsection{A Step in the Computational Frame}
\label{subsec:Jlm_tensor}

A convenient way to derive the $J_{lm}^A(fD, \hat{k})$'s is to first point $\hat{k}$ to the $\hat{z}$ direction, and then at the end perform an arbitrary rotation \cite{Liu:2022skj}. This implies the simplification
\begin{equation}
    Y_{LM}\left(\hat{k}=\hat{z}\right) \equiv \sqrt{\dfrac{2L+1}{4\pi}} \delta_{M0} \,.
\end{equation}
Furthermore, it is useful to proceed using a right- and left-handed complex circular polarization basis,
\begin{equation}
    \varepsilon^\text{R} = \dfrac{\varepsilon^+ + i \varepsilon^\times}{\sqrt{2}} \ \ \ \ \ \text{and} \ \ \ \ \varepsilon^\text{L} = \dfrac{\varepsilon^+ - i \varepsilon^\times}{\sqrt{2}} \,.
\end{equation}
In this case, the contraction of the `detector tensor', $\left( \hat{e}^i \otimes \hat{e}^j \right)$, with the polarization basis tensors give
\begin{equation}
    \left( \hat{e}^i \otimes \hat{e}^j \right) \varepsilon_{ij}^\text{R, L} = \sqrt{\dfrac{16\pi}{15}} Y_{2 \pm 2} \left( \hat{e} \right) \,,
\end{equation}
where the helicity R (L) takes on the index $m = + 2$ ($-2$).

Substituting the above result into Eq. \eqref{eq:Jlm_def}, we obtain
\begin{equation}
    J_{lm}^\text{R,L} \left( fD, \hat{z} \right) = 4 \pi \int_0^{2\pi f D} dx \ e^{i x} \sum_{L} i^L \sqrt{\dfrac{2L+1}{15}} j_L(x) \int_{S^2} d\hat{e} \ Y_{2\pm2}\left(\hat{e}\right) Y_{L0}\left( \hat{e} \right) Y_{lm}^*\left(\hat{e}\right) \,.
\end{equation}
The spherical harmonics integral vanishes unless $m = \pm 2$ and $L = l - 2, l, l + 2$. In the nonvanishing cases, the integral can be shown to be
\begin{eqnarray}
\int_{S^2} d\hat{e} \ Y_{2\pm 2}\left(\hat{e}\right) Y_{(l - 2) 0}\left(\hat{e}\right) Y^*_{l \pm 2}\left(\hat{e}\right) &=& \sqrt{\dfrac{15}{32\pi}} \left( \dfrac{(l-1)l(l+1)(l+2)}{(2l-3)(2l-1)^2(2l+1)} \right)^{1/2} \,, \nonumber \\
&& \nonumber \\ 
\int_{S^2} d\hat{e} \ Y_{2\pm 2}\left(\hat{e}\right) Y_{l0}\left(\hat{e}\right) Y^*_{l \pm 2}\left(\hat{e}\right) &=& - \sqrt{\dfrac{15}{8\pi}} \left( \dfrac{(l-1)l(l+1)(l+2)}{(2l-1)^2(2l+3)^3} \right)^{1/2} \,, \nonumber \\
&& \nonumber \\
\int_{S^2} d\hat{e} \ Y_{2\pm 2}\left(\hat{e}\right) Y_{(l + 2) 0}\left(\hat{e}\right) Y^*_{l \pm 2}\left(\hat{e}\right) &=& \sqrt{\dfrac{15}{32\pi}} \left( \dfrac{
(l-1)l(l+1)(l+2)}{(2l+1)(2l+3)^2(2l+5)} \right)^{1/2} \nonumber \,. \\
&& \label{eq:tripleSintegral}
\end{eqnarray}
Using the identities \eqref{eq:tripleSintegral}, we directly obtain
\begin{equation}
\begin{split}
    J_{lm}^\text{R,L}\left(fD, \hat{z}\right) = &-\delta_{m\pm 2} 2 \pi i^l \sqrt{ \dfrac{2l+1}{8\pi} \dfrac{(l + 2)!}{(l - 2)!} } \int_0^{2\pi fD} dx \ e^{ix} \\
    & \ \ \times \left( \dfrac{j_{l-2}(x)}{(2l-1)(2l+1)} + \dfrac{2j_l(x)}{(2l-1)(2l+3)} + \dfrac{j_{l+2}(x)}{(2l+1)(2l+3)} \right) \,,
\end{split}
\end{equation}
which with a recursion relation,
\begin{equation}
\label{eq:bessel_id1}
    \dfrac{j_l(x)}{x} = \dfrac{j_{l-1}(x) + j_{l+1}(x)}{2l+1} \,,
\end{equation}
we are able to simplify into
\begin{equation}
    J_{lm}^\text{R,L}\left(fD, \hat{z}\right) = - \delta_{m\pm 2} \sqrt{\dfrac{2l+1}{4\pi}} \left( \sqrt{2} \pi i^l \sqrt{ \dfrac{(l + 2)!}{(l - 2)!} } \int_0^{2\pi fD} dx \ e^{ix} \dfrac{j_l(x)}{x^2} \right) \,.
\end{equation}
In this last expression, the factor $\sqrt{(2l + 1)/4\pi}$ corresponds to an arbitrary rotational degree of freedom. For an isotropic SGWB, the important quantity is enclosed in the parenthesis in the above result. 

To generalize the result, we reorient $\hat{z}$ into an arbitrary direction, $\hat{k} = (\theta,\phi)$,
\begin{equation}
    J^A_{lm}\left( fD, \hat{k} \right) = \sum_{m'} D_{m'm}^{l*} \left(-\alpha,-\theta,-\phi\right) J^A_{lm'}\left( fD, \hat{z} \right) \,,
\end{equation}
where $D_{m'm}^l(-\alpha,-\theta,-\phi)$ is the Wigner matrix given by
\begin{equation}
    D_{m'm}^l(-\alpha,-\theta,-\phi) = \sqrt{\dfrac{4\pi}{2l + 1}} \, _{-m'}Y_{lm}\left(\theta,\phi\right) e^{i m' \alpha} \,.
\end{equation}
The $\,_s Y_{lm}\left(\hat{e}\right)$'s  are known as spin-weighted spherical harmonics; the standard $Y_{lm}(\hat{e})$ corresponds to the spinless case $s=0$. Following this procedure, we obtain our desired final result
\begin{equation}
\label{eq:Jlm_tensor}
    J_{lm}^\text{R,L}\left(fD, \hat{k}\right) = - _{\mp 2}Y_{lm}^* \left( \hat{k} \right) e^{\mp 2i \alpha} \left( \sqrt{2}\pi i^l \sqrt{ \dfrac{(l + 2)!}{(l - 2)!} } \int_0^{2\pi fDv} \dfrac{dx}{v} \ e^{ix/v} \dfrac{j_l(x)}{x^2} \right) \,,
\end{equation}
where the upper (lower) signs belong to R (L). The factor $e^{i m' \alpha}$ is a redundant phase associated to a remaining rotational degree of freedom about the $\hat{k}$ axis. This drops out for the relevant PTA observables.

\subsection{The ORF and the Angular Power Spectrum}
\label{subsec:power_spectra_tensor}

By substituting Eq. \eqref{eq:Jlm_tensor} into Eq. \eqref{eq:orf_general}, using the spherical harmonics addition theorem,
\begin{equation}
\label{eq:addition_theorem}
    P_l \left( \hat{e}_a \cdot \hat{e}_b \right) = \dfrac{4\pi}{2l+1} \sum_{m} Y_{lm}\left(\hat{e}_a\right) Y_{lm}^*\left(\hat{e}_b\right) \,,
\end{equation}
and accounting for the equal contributions of the right- and left-handed helicity terms, we obtain the spatial correlation
\begin{equation}
\label{eq:gamma_harmonic_series}
\begin{split}
    \gamma\left( f D_a, fD_b, \hat{e}_a \cdot \hat{e}_b \right) = & \sum_{A=+,\times}\gamma^A(fD_a, fD_b, \hat{e}_a\cdot\hat{e}_b) \\
    = & \ \sum_l \dfrac{2l + 1}{4\pi} C_l \left( f D_a, f D_b \right) P_l \left( \hat{e}_a \cdot \hat{e}_b \right) \,,
\end{split}
\end{equation}
where the tensor (T) angular power spectrum multipoles are given by
\begin{equation}
\label{eq:cls_tensor}
    C_l^\text{T}\left( f D_a, f D_b \right) = \dfrac{J^\text{T}_l\left(f D_a\right) J^{\text{T}*}_l\left(f D_b\right)}{\sqrt{\pi}} \,,
\end{equation}
and
\begin{equation}
\label{eq:jls_tensor}
    J^\text{T}_l\left(fD\right) = \sqrt{2} \pi i^l \sqrt{\dfrac{(l + 2)!}{(l - 2)!}} \int_0^{2\pi fD} dx \ e^{ix} \dfrac{j_l(x)}{x^2} \,.
\end{equation}

In the limit, $fD \gg 1$, relevant to PTAs, it is worth noting that the integral can to a good approximation be written as
\begin{equation}
    \int_0^\infty dx \ e^{ix} \dfrac{j_l(x)}{x^2} = 2 i^{l-1} \dfrac{(l - 2)!}{(l + 2)!} \,.
\end{equation}
In this case, the angular power spectrum reduces to the HD correlation given by
\begin{equation}
\label{eq:cls_hd}
    C_l^\text{T}\left( f D_a, f D_b \right)\sim C_l^{\rm HD}=\dfrac{8\pi^{3/2}}{(l + 2)(l + 1)l(l - 1)} \,.
\end{equation}
Note that for this limit the frequency and pulsar distance dependencies completely drop out. The same result can be obtained by dropping the pulsar term in the standard configuration space derivation. Substituting the HD correlation power spectrum into Eq. \eqref{eq:gamma_harmonic_series} and closing the sum, we recover the HD curve \cite{Hellings:1983fr},
\begin{equation}
\label{eq:hd_gamma}
\gamma\left( f D_a, fD_b, \hat{e}_a \cdot \hat{e}_b \right) \sim \gamma^{\rm HD}(\hat{e}_a \cdot \hat{e}_b)= \overline{\cal N} \left( {\mathbf{\Gamma}}(\hat{e}_a \cdot \hat{e}_b) + \dfrac{\delta_{ab}}{2} \right) \,,
\end{equation}
where $\overline{\cal N}$ is a constant and ${\mathbf{\Gamma}}(x)$ is
\begin{equation}
\label{eq:hd_curve}
    {\mathbf{\Gamma}}(x) = \dfrac{1}{2} - \dfrac{1}{4} \left( \dfrac{1-x}{2} \right) + \dfrac{3}{2} \left( \dfrac{1-x}{2} \right) \ln \left( \dfrac{1-x}{2} \right) \,.
\end{equation}
This is the main result of this section.

It is worth noting that the constant $\overline{\cal N}$ is an arbitrary normalization \cite{Romano:2023zhb}. In the original derivation by Hellings and Downs \cite{NANOGrav:2023hde}, this is given by $\overline{\cal N}=2/3$, turning in a `zero lag' ${\mathbf{\Gamma}(1)}=1/3$, whereas in today's standard practice, this is fixed to unity, ${\mathbf{\Gamma}(1)}=1/2$. Most importantly, in the HD limit, a term $\delta_{ab}/2$ is accounted for by hand that correspond to the power at small scales, owed to the pulsar term. The correlation at large angular scales can thus be fully accounted for by the Earth term, while at small scales is half and half between the Earth and the pulsar terms.

\begin{figure}[h!]
    \centering
    \includegraphics[width=0.975\linewidth]{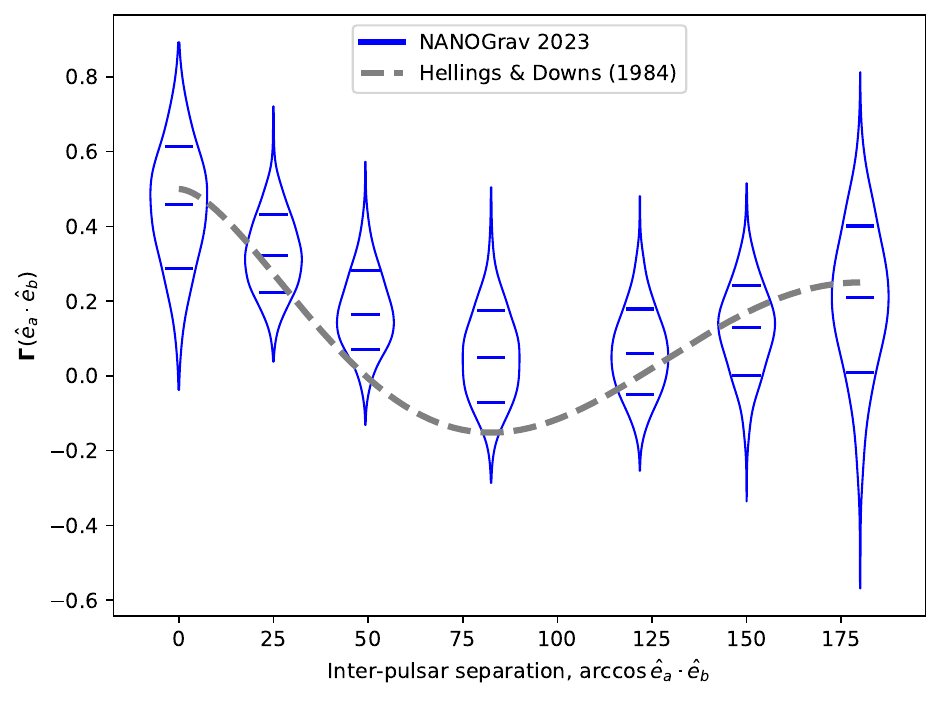}
    \caption{The overlap reduction function in the NANOGrav 15-year data set \cite{NANOGrav:2023gor} (blue violins) and the Hellings and Downs curve (gray dashed line).}
    \label{fig:orf}
\end{figure}

Figure \ref{fig:orf} shows the measured ORF, $\mathbf{\Gamma}(\hat{e}_a\cdot \hat{e}_b)$, between pulsar pairs in a PTA and the HD curve. The finite distance effects can also be computed using the full Eqs. (\ref{eq:gamma_harmonic_series}-\ref{eq:jls_tensor}), noting that the upper limit of the integral depends on $fD \sim 10^2 \times (f/{1 \ {\rm nHz}}) \times (D/{1 \ {\rm kpc}})$. These corrections are relevant at small angles, i.e., sub-degrees \cite{Ng:2021waj}, and are beyond the sensitivity of PTAs given their present sky coverages. As we shall show, the uncertainty in the present PTA measurements of the correlation opens up the interpretation of the gravitational aspect of the signal to theoretical uncertainties, non-Gaussianity, alternative theories of gravity, and anisotopy and polarization.

\section{Beyond the Hellings and Downs Curve}
\label{sec:beyond_hd_curve}

In this section, we focus on theoretical progress that has been made recently, and which foreshadows PTA science's future milestones. We particularly discuss the cosmic variance (Section \ref{subsec:cosmic_variance}), the cumulants of the one- and two-point functions of the SGWB (Section \ref{subsec:cumulants}), subluminal GWs (Section \ref{subsec:subluminal_gws}), and anisotropy and polarization (Section \ref{subsec:anisotropy}).

\subsection{Cosmic Variance}
\label{subsec:cosmic_variance}

Recently, theoretical uncertainties were realized in the context of PTAs. The most notable of this is the cosmic variance (Eq. \eqref{eq:cv_ps}), an uncertainty tied to interfering sources of GW in a PTA \cite{Allen:2022dzg, Allen:2022ksj, Bernardo:2022xzl, Bernardo:2023bqx, Caliskan:2023cqm, Allen:2024rqk, Allen:2024uqs, Allen:2024bnk, Wu:2024xkp}. Another widely adopted interpretation of the cosmic variance is an uncertainty due to the fact that cosmological measurements can be done in only a single universe. In this scenario, the cosmic variance can be referred to as an ensemble variance.

In this section, we briefly review the cosmic variance in PTA using the power spectrum/harmonic space formalism that has been utilized throughout this review.

\subsubsection{Full-sky averages and the correlation operator}

We consider that there is a large number of pulsars on the sky and perform full-sky averages, $\{ \cdots \}_{\zeta}$, with a fixed angular separation $\zeta$ between pulsar pairs. In symbols, we denote a full-sky average quantity with curly brackets,
\begin{equation}
    \{ \cdots \}_\zeta = \int d\Omega d\Omega' \cdots d\Omega'' \left( \cdots \right) \,.
\end{equation}
For a product of spherical harmonics, the full-sky average integral can be shown to be \cite{Ng:1998eok, Ng:1997ez}
\begin{equation}
    \{ Y_{l'm'}^*\left(\hat{n}'\right) Y_{lm}\left(\hat{n}\right) \}_\zeta = P_l \left( \cos \zeta \right) \dfrac{\delta_{ll'} \delta_{mm'}}{4\pi} \,.
\end{equation}

To introduce this operation, we shall show that the full-sky average of the correlation through a `correlation operator' leads to the HD curve.

It is useful to associate the GW correlation with a correlation operator,
\begin{equation}
\label{eq:cop}
    {\bm\gamma}_{ab} = {\bm\beta}_a {\bm\beta}_b \,,
\end{equation}
where ${\bm \beta}_a$ admits a harmonic series representation,
\begin{equation}
\label{eq:beta_hseries}
    {\bm \beta}_a = \sum_{lm} {\bm b}_{lm} Y_{lm}\left(\hat{e}_a\right) \,.
\end{equation}
Note that while we have dropped explicit dependencies on the time, frequency, and distance to pulsars for brevity, the formalism remains general. The abstract quantities ${\bm \gamma}_{ab}$ and ${\bm \beta}_a$ represent the timing residual cross correlation, $\langle {\bm r}_a {\bm r}_b \rangle$, and the pulsar timing residual, ${\bm r}_a$, respectively, and carry their physical meaning. This connection with an observable implies that ${\bm \beta}_a^\dagger = {\bm \beta}_a$, and by extension ${\bm b}^\dagger_{lm} = (-1)^m{\bm b}_{l-m}$. 

To establish the connection more clearly, we show that the correlation can be obtained by taking the ensemble average of the correlation operator. We substitute \eqref{eq:beta_hseries} into \eqref{eq:cop}, and take the ensemble average to obtain
\begin{equation}
    \langle {\bm\gamma}_{ab} \rangle = \sum_{l_1 m_1} \sum_{l_2 m_2} \langle {\bm b}_{l_1 m_1} {\bm b}_{l_2 m_2} \rangle Y_{l_1 m_1} \left( \hat{e}_a \right) Y_{l_2 m_2} \left( \hat{e}_b \right) \,.
\end{equation}
On the other hand, spatial isotropy implies that
\begin{equation}
\label{eq:isotropic}
    \langle {\bm b}^\dagger_{l_1 m_1} {\bm b}_{l_2 m_2} \rangle = \overline{C}_{l_1} \delta_{l_1 l_2} \delta_{m_1 m_2} \,,
\end{equation}
or equivalently,
\begin{equation}
\label{eq:isotropic_real}
    \langle {\bm b}_{l_1 m_1} {\bm b}_{l_2 m_2} \rangle = (-1)^{m_1} \overline{C}_{l_1} \delta_{l_1 l_2} \delta_{m_1 -m_2} \,,
\end{equation}
for some $\overline{C}_l$'s. The ensemble average of the correlation operator thus becomes
\begin{equation}
\label{eq:2-beta_ensave}
    \langle {\bm\gamma}_{ab} \rangle = \sum_{l} \dfrac{2l + 1}{4\pi} \overline{C}_l P_l \left( \hat{e}_a \cdot \hat{e}_b \right) \,,
\end{equation}
where we have utilized the addition theorem \eqref{eq:addition_theorem}. This result ties the correlation with the correlation operator through the ensemble average,
\begin{equation}
    \gamma\left(fD_a, fD_b,  \hat{e}_a\cdot\hat{e}_b\right) = \langle {\bm\gamma}_{ab} \rangle \,,
\end{equation}
where now the $C_l$'s can be identified to be the angular power spectrum multipoles of the SGWB, e.g., Eq. \eqref{eq:cls_tensor} or \eqref{eq:cls_hd}.

Now, we derive the full-sky average of the correlation operator,
\begin{equation}
\begin{split}
    \{ {\bm\gamma}_{ab} \}_\zeta = & \sum_{l_1 m_1} \sum_{l_2 m_2} {\bm b}_{l_1m_1} {\bm b}_{l_2m_2} \{ Y_{l_1m_!}\left( \hat{e}_a \right) Y_{l_2m_2}\left(\hat{e}_b\right) \}_\zeta \\
    = & \sum_{l_1 m_1} \sum_{l_2 m_2} {\bm b}_{l_1m_1} {\bm b}_{l_2m_2} \left( (-1)^{m_1} P_{l_1} \left( \cos \zeta \right) \dfrac{\delta_{l_1l_2} \delta_{m_1m_2}}{4\pi} \right)\,, \\
\end{split}
\end{equation}
where we reiterate that $\zeta$ is a fixed angular separation between pulsar pairs. In the last line, let us define an `angular power spectrum operator'
\begin{equation}
\label{eq:ps_operator}
    {\bm C}_l = \sum_m \dfrac{{\bm b}_{lm}^\dagger {\bm b}_{lm}}{2l + 1} \,.
\end{equation}
It is easy to show that $\langle {\bm C}_l \rangle =  C_l$, hence the name. Then, the full-sky average of the correlation operator finally becomes
\begin{equation}
\label{eq:fsave_hd}
    \{ {\bm\gamma}_{ab} \}_\zeta =
    \sum_{l} \dfrac{ 2l + 1 }{4\pi} {\bm C}_l P_l \left( \cos\zeta \right) \,.
\end{equation}
Clearly, the ensemble average gives the correlation,
\begin{equation}
    \gamma\left(fD_a, fD_b, \hat{e}_a\cdot\hat{e}_b=\cos\zeta\right) \equiv \langle \{ {\bm\gamma}_{ab} \}_\zeta \rangle \,.
\end{equation}

\subsubsection{The cosmic variance as a sky average}

The cosmic variance of the correlation, ${\rm CV}\left[ \{ {\bm \gamma}_{ab} \}_\zeta \right]$, can be obtained by evaluating the full-sky and ensemble averages in
\begin{equation}
\label{eq:cv_fullskydef}
    \text{CV}\left[ \{ {\bm \gamma}_{ab} \}_\zeta \right] = \langle \{ {\bm\gamma}_{ab} \}_\zeta^2 \rangle - \langle \{ {\bm\gamma}_{ab} \}_\zeta \rangle ^2 \,.
\end{equation}
Note that the second term is the square of the mean.

To deal with the first term, we square the full-sky averaged correlation operator,
\begin{equation}
\begin{split}
    \{ {\bm\gamma}_{ab} \}_\zeta^2 = & \left( \sum_{l} \dfrac{2l+1}{4\pi} {\bm C}_l P_l (\cos\zeta) \right)^2 \\
    = & \sum_{ll'} \dfrac{(2l + 1)(2l' + 1)}{(4\pi)^2} {\bm C}_l {\bm C}_{l'} P_l \left( \cos \zeta \right) P_{l'} \left( \cos \zeta \right) \,,
\end{split}
\end{equation}
and evaluate its ensemble average,
\begin{equation}
    \langle \{ {\bm\gamma}_{ab} \}_\zeta^2 \rangle = \sum_{ll'} \dfrac{(2l + 1)(2l' + 1)}{(4\pi)^2} \langle {\bm C}_l {\bm C}_{l'} \rangle P_l \left( \cos \zeta \right) P_{l'} \left( \cos \zeta \right) \,.
\end{equation}
Then, using the definition of the angular power spectrum operator (Eq. \ref{eq:ps_operator}), we write down
\begin{equation}
\begin{split}
    \langle {\bm C}_l {\bm C}_{l'} \rangle = & \bigg \langle \sum_m \dfrac{{\bm b}_{lm}^\dagger {\bm b}_{lm}}{2l + 1} \sum_{m'} \dfrac{{\bm b}_{l'm'}^\dagger {\bm b}_{l'm'}}{2l' + 1} \bigg \rangle \\
    = & \sum_{mm'} \dfrac{1}{(2l+1)(2l'+1)} \langle {\bm b}_{lm}^\dagger {\bm b}_{lm} {\bm b}_{l'm'}^\dagger {\bm b}_{l'm'} \rangle \,.
\end{split}
\end{equation}
Furthermore, because the SGWB is Gaussian, by definition, the four-point function factorizes into a product-sum of two-point functions,
\begin{equation}
\label{eq:wickrotation}
\begin{split}
    \langle {\bm b}^\dagger_{l_1 m_1} {\bm b}_{l_2 m_2} {\bm b}^\dagger_{l_3 m_3} {\bm b}_{l_4 m_4} \rangle = \ &  C_{l_1} C_{l_3} \delta_{l_1 l_2} \delta_{m_1 m_2} \delta_{l_3 l_4} \delta_{m_3 m_4} + C_{l_1} C_{l_2} \delta_{l_1 l_4} \delta_{m_1 m_4} \delta_{l_2 l_3} \delta_{m_2 m_3} \\
    & + (-1)^{m_1} (-1)^{m_2} C_{l_1} C_{l_2} \delta_{l_1 l_3} \delta_{m_1 -m_3} \delta_{l_2 l_4} \delta_{m_2 -m_4} \,.
\end{split}
\end{equation}
This implies that
\begin{equation}
    \langle {\bm b}_{lm}^\dagger {\bm b}_{lm} {\bm b}_{l'm'}^\dagger {\bm b}_{l'm'} \rangle = C_l C_{l'} + C_l^2 \delta_{ll'} \delta_{mm'} + C_l^2 \delta_{ll'} \delta_{m-m'} \,,
\end{equation}
which we use to simplify $\langle {\bm C}_l {\bm C}_{l'} \rangle$. By using the above expressions, summing over the Kronecker deltas, we obtain
\begin{equation}
\label{eq:secondmoment_corr_skyave}
    \langle \{ {\bm\gamma}_{ab} \}_\zeta^2 \rangle = \left( \sum_{l} \dfrac{2l + 1}{4\pi} C_l P_l \left( \cos \zeta \right) \right)^2 + \sum_l \dfrac{2l + 1}{8\pi^2} C_l^2 P_l\left(\cos\zeta\right)^2 \,.
\end{equation}
The first term above, $\langle \{ {\bm \beta}_a^\dagger \bm{\beta}_b \}_\zeta \rangle ^2 = \gamma_{ab}(\zeta)^2$, which is equal to the square of the correlation. Therefore, substituting this into Eq. \eqref{eq:cv_fullskydef}, we obtain the cosmic variance of the correlation \cite{Allen:2022dzg, Bernardo:2022xzl},
\begin{equation}
\label{eq:cv_ps}
\text{CV}\left[ \{ {\bm \gamma}_{ab} \}_\zeta \right] = \sum_l \dfrac{2l + 1}{8\pi^2} C_l^2 P_l\left(\cos\zeta\right)^2 \,.
\end{equation}
The above important expression was first derived in Ref. \citen{Allen:2022dzg} (Appendix C, Eq. (C53)) in GR, and independently in Ref. \citen{Bernardo:2022xzl} in the context of non-Einsteinian and subluminal GWs.

\begin{figure}[h!]
    \centering
    \includegraphics[width=0.975\textwidth]{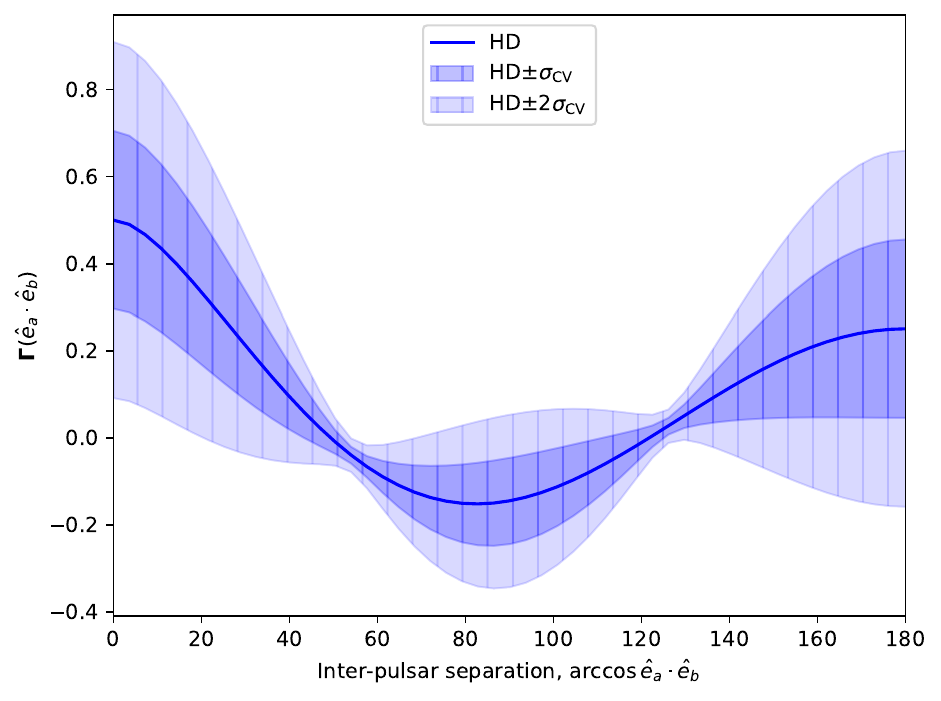}
    \caption{The Hellings and Downs curve \cite{Hellings:1983fr} and its cosmic variance \cite{Allen:2022dzg, Bernardo:2022xzl}.}
    \label{fig:cosmic_variance}
\end{figure}

This is the limit of having a sufficiently large number of pulsars to pair up to measure the correlation, and agrees quite well with Ref. \citen{Allen:2022dzg}. Figure \ref{fig:cosmic_variance} shows the cosmic variance (Eq. \eqref{eq:cv_ps}) of the Hellings and Downs correlation (Eq. \eqref{eq:cls_hd}).

The impact of the cosmic variance has been considered in the search for the SGWB including non-Einsteinian GW polarizations in Refs. \citen{Bernardo:2023bqx, Bernardo:2023zna}, highlighting the significance of the first two leading moments of a distribution (instead of only the mean) to look for a signal. Recently, its impact was revisited in the NANOGrav 15-yr data set in Ref. \citen{Wu:2024xkp}, showing a huge increase in the evidence for the SGWB detection. When testing gravity, the impact of the cosmic variance cannot be understated \cite{Bernardo:2022xzl}.

The variance of the angular power spectrum can also be computed, leading to a friendly expression,
\begin{equation}
\label{eq:variance_cls}
    \dfrac{\Delta C_l}{ C_l } = \sqrt{\dfrac{2}{2l+1}} \,.
\end{equation}
A detailed calculation is given in Ref. \citen{Bernardo:2022xzl}. This is a similar equation for the temperature anisotropies of the cosmic microwave background (CMB), and implies that increasing the angular resolution allows measurements of the angular power spectrum with a larger $l$, e.g., a $\Delta \zeta  = 0.1^\circ$ resolution will let an experiment probe up to, $l \leq 180^\circ/\Delta \zeta \sim 1800$, with a variance $\Delta C_l/C_l \sim 1/\sqrt{1800} \sim 1/42$. In Section \ref{subsec:subluminal_gws}, we shall find that noise-free or cosmic-variance limited measurements of the angular power spectrum are most suitable to constrain nanohertz gravity \cite{Bernardo:2023pwt}.

\subsection{Cumulants of the One- and Two-Point Functions}
\label{subsec:cumulants}

The Gaussianity of a signal offers insights into its origin, and can be probed through the statistical properties of the relevant observables. We review the foundations of Gaussian statistics (Section \ref{subsubsec:gaussian_statistics}), and how the signatures of Gaussianity in the one- and two-point functions of a SGWB can be constrained using PTA simulations (Sections \ref{subsubsec:one_point}-\ref{subsubsec:two_point}). The methods discussed are also applicable to SGWB in other regimes, in the millihertz and sub-kilohertz GW bands.

\subsubsection{Gaussian statistics}
\label{subsubsec:gaussian_statistics}

A Gaussian stochastic field, $\Psi(x)$, is completely specified by a two-point function, $\langle \Psi(x) \Psi(x') \rangle$ \cite{Weinberg:2008zzc}, with odd moments vanishing,
\begin{equation}
\label{eq:odd_moments_gaussian}
    \langle \Psi(x) \rangle = \langle \Psi(x) \Psi(x') \Psi(x'') \rangle = \cdots = 0 \,,
\end{equation}
and even moments factorizing into a product-sum of two-point functions, $\langle \Psi(x) \Psi(x') \rangle$. The four-point function becomes \cite{Isserlis_Theorem}
\begin{equation}
\label{eq:4p_gaussian}
\begin{split}
    & \langle \Psi_1 \Psi_2 \Psi_3 \Psi_4 \rangle 
    = \langle \Psi_1 \Psi_2 \rangle \langle \Psi_3 \Psi_4 \rangle + \langle \Psi_1 \Psi_3 \rangle \langle \Psi_2 \Psi_4 \rangle + \langle \Psi_1 \Psi_4 \rangle \langle \Psi_2 \Psi_3 \rangle \,.
\end{split}
\end{equation}
The notation $\Psi_i = \Psi(x_i)$ is used for brevity. The six-point, eight-point, and higher-point functions follow suit, including \cite{Srednicki:1993ix, Gangui:1994wh}
\begin{equation}
\label{eq:6p_gaussian}
    \langle \Psi_1^3 \Psi_2^3 \rangle = 9 \langle \Psi_1^2 \rangle \langle \Psi_2^2 \rangle \langle \Psi_1 \Psi_2 \rangle + 6 \langle \Psi_1 \Psi_2 \rangle^3 \,,
\end{equation}
and
\begin{equation}
\label{eq:8p_gaussian}
    \langle \Psi_1^4 \Psi_2^4 \rangle = 9 \langle \Psi_1^2 \rangle^2 \langle \Psi_2^2 \rangle^2 + 72 \langle \Psi_1^2 \rangle \langle \Psi_2^2 \rangle \langle \Psi_1 \Psi_2 \rangle^2 + 24 \langle \Psi_1 \Psi_2 \rangle^4 \,.
\end{equation}

In a Gaussian field, all information in the higher-points and cumulants are packed into the two-point function. Thus, the Gaussianity of a field can be tested by measuring the higher-point functions and their cumulants to look for redundancy beyond that already given by the lowest moment/mean.

In practice, an observation $O(z)$ can be taken as ${\cal G}\left[ \Psi(x) \right]$ where ${\cal G}$ is a functional of the field $\Psi(x)$. If $\Psi(x)$ is a Gaussian field, and ${\cal G}$ is a linear functional, then the observation $O(z)$ can be expected to similarly be Gaussian; that is, the statistics of $O(z)$ will be completely specified by a two-point function $\langle O(z) O(z') \rangle$, and its higher moments will factorize similarly as in Eqs. (\ref{eq:odd_moments_gaussian}-\ref{eq:8p_gaussian}). This is put to test in CMB analysis to constrain potential non-Gaussianities in the data \cite{Planck:2019evm}, where $\Psi(x) \equiv \delta (x)$ are the density perturbations, and $O(z) \equiv \Delta T(z)/{\overline T}$ is the temperature fluctuation. Analogously, in PTA science, a similar test can be proposed, 
where $\Psi(x) \equiv h(x)$ are GWs, and $O(z) \equiv r(z)$ are the pulsar timing residuals.

Practical considerations set apart CMB observations and SGWB measurements in PTA. For the CMB, the presence of power at all scales and subdominant instrumental noise let one-point statistics to be usable to test the Gaussianity of the signal. For PTAs, neither conditions are satisfied; loud intrinsic pulsar noises are generally present and the power is concentrated in large scale modes, particularly in the quadrupole. The implication of this is discussed in detail in Ref. \citen{Bernardo:2024uiq}.

\subsubsection{One-Point Cumulants}
\label{subsubsec:one_point}

Consider an isotropic and Gaussian SGWB, with a power spectrum given by Eq. \eqref{eq:gwb_power_spectrum}. Utilizing Eqs. (\ref{eq:odd_moments_gaussian}-\ref{eq:8p_gaussian}), it can be shown that the mean $\mu$, variance $\sigma$, skewness ${\cal S}$, and kurtosis $\cal K$ of pulsar timing residuals are given by
\begin{equation}
    \dfrac{\mu}{\sqrt{\gamma_{aa}}} = 0 \pm \sqrt{ \dfrac{1}{2\gamma_{aa}} \int_{-1}^1 dx_{ab} \ \gamma\left( x_{ab} \right) } \,,
\end{equation}
\begin{equation}
    \dfrac{\sigma^2}{\gamma_{aa}} = 1 \pm \sqrt{ \dfrac{1}{\gamma_{aa}^2} \int_{-1}^1 dx_{ab} \ \gamma\left( x_{ab} \right)^2 } \,,
\end{equation}
\begin{equation}
    {\cal S} = 0 \pm \sqrt{ \dfrac{3}{\gamma_{aa}^3} \int_{-1}^1 dx_{ab} \ \gamma\left( x_{ab} \right)^3 } \,,
\end{equation}
and 
\begin{equation}
    {\cal K} = 3 \pm \sqrt{ \dfrac{36}{\gamma_{aa}^2} \int_{-1}^1 dx_{ab} \ \gamma\left( x_{ab} \right)^2 + \dfrac{12}{\gamma_{aa}^4} \int_{-1}^1 dx_{ab} \ \gamma\left( x_{ab} \right)^4 } \,,
\end{equation}
respectively, where $\gamma(x_{ab})$ is the correlation, i.e., Eq. \eqref{eq:gamma_harmonic_series}, and $\gamma_{aa}=\gamma(0)$ is the auto-correlation. The integral over $x_{ab}$ is akin to a full-sky averaging over pulsar pairs, and the quantity in the square roots are the cosmic variances of mean, variance, skewness, and kurtosis which can be derived straightforwardly using Gaussian combinatorics \cite{Srednicki:1993ix, Gangui:1994wh, Bernardo:2024uiq}. For the HD correlation, this gives a skewness ${\cal S}\sim {\cal S}^{\rm HD}=0 \pm 0.133$ and a kurtosis ${\cal K}\sim {\cal K}^{\rm HD}=3 \pm 1.24$.

However, the above one-point measures of a SGWB's Gaussianity will be completely obscured by pulsar red noises (Section \ref{subsubsec:stochastic_part}) which dominate one-point statistics. It is also worth noting that the distribution of the sample statistics do not strictly follow a Gaussian probability distribution function because of the leading quadrupole. Nonetheless, the cosmic variances above can be confirmed by numerically simulating the ensemble statistics' non-Gaussian probability distribution function, $P(X)$, and computing the variance using $\langle X^2 \rangle \equiv\int dX \ P(X) X^2$ where the random variable $X$ represents the sample mean, variance, skewness and kurtosis \cite{Bernardo:2024uiq}.

\subsubsection{Two-Point Cumulants}
\label{subsubsec:two_point}

In contrast with one-point statistics, two-point statistics, now accounting for the sky coverage of a PTA, turns out to be a more promising direction to probe the Gaussianity of the SGWB, because the red noise (Section \ref{subsubsec:stochastic_part}) in pulsars are uncorrelated over pulsar pairs.

\begin{figure}[h!]
    \centering
    \includegraphics[width=1.0\textwidth]{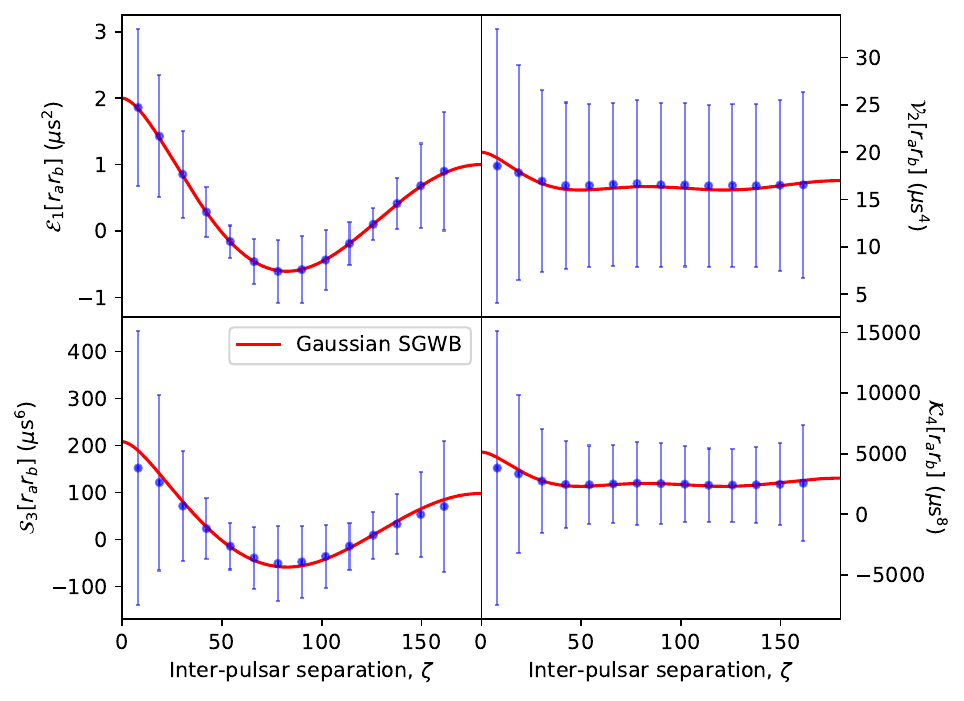}
    \caption{Mean (top left), variance (top right), skewness (lower left), and kurtosis (lower right) of the two-point function, in the 1st $\alpha$-bin (Eq. \eqref{eq:residual_fourier_series}), in a noise-free PTA simulation (blue points and error bars/cosmic variance). This is constructed by repeatedly computing the sample statistics with 100 pulsars randomly distributed over the sphere that are correlated by a SGWB ($A_{\rm gw}\sim 10^{-15}$ and $\gamma_{\rm gw}=13/3$). The red curves are given by Eqs. (\ref{eq:rarb_sgwb_mean}-\ref{eq:rarb_sgwb_kurtosis}).}
    \label{fig:pulsar_2p}
\end{figure}

Once again, assuming an isotropic and Gaussian SGWB, and using Gaussian combinatorics (Eqs. (\ref{eq:odd_moments_gaussian}-\ref{eq:8p_gaussian})), it can be shown that the mean ${\cal E}_1$, variance ${\cal V}_2$, skewness ${\cal S}_3$ and kurtosis ${\cal K}_4$ of the timing residual cross correlation are given by \cite{Bernardo:2024uiq}
\begin{eqnarray}
\label{eq:rarb_sgwb_mean}
    {\cal E}_1[ r_a r_b ] &=& \tilde{a}^2 \gamma( \hat{e}_a \cdot \hat{e}_b ) \,, \\
\label{eq:rarb_sgwb_variance}
    {\cal V}_2[r_a r_b] &=& \tilde{a}^4 \left( 1+\gamma( \hat{e}_a \cdot \hat{e}_b )^2 \right) \,, \\
\label{eq:rarb_sgwb_skewness}
    {\cal S}_3[r_a r_b] &=& 2 \tilde{a}^6 \gamma( \hat{e}_a \cdot \hat{e}_b ) \left( 3 + \gamma( \hat{e}_a \cdot \hat{e}_b )^2 \right) \,, \\
\label{eq:rarb_sgwb_kurtosis}
    {\cal K}_4[r_a r_b] &=& 3 \tilde{a}^8 \left( 3 + 14 \gamma( \hat{e}_a \cdot \hat{e}_b )^2 + 3 \gamma( \hat{e}_a \cdot \hat{e}_b )^4 \right) \,,
\end{eqnarray}
where ${\tilde a}^2$ is a constant proportional to the power spectrum at a given frequency, and $\gamma( \hat{e}_a \cdot \hat{e}_b )$ is the ORF/correlation between pulsar pairs, i.e., Eq. \eqref{eq:gamma_harmonic_series}. These equations manifest the sentiment that all the higher cumulants of the two-point function reduce to products of the two-point correlation function $\gamma( \hat{e}_a \cdot \hat{e}_b )$, by virtue of Gaussianity. For this case, Gaussianity implies that the cumulants of the timing residual correlation in a PTA must give no information beyond the HD curve \cite{Planck:2019evm}. Observational departures from Eqs. (\ref{eq:rarb_sgwb_mean}-\ref{eq:rarb_sgwb_kurtosis}) will be indicative of non-Gaussianity of a SGWB. We emphasize that the cosmic variance is not considered in the above expressions, which requires a computation up to a sixteen-point function. The signal of a SGWB including the cosmic variance in the cumulants of the two-point function can instead be simulated for any PTA setup, as shown in Figure \ref{fig:pulsar_2p}.

The result shows the signature of Gaussianity of a SGWB through the cumulants of the two-point function, in this case, in a PTA simulation. Perhaps worth highlighting is that the skewness also reduces to the Hellings and Downs curve, as explained by Eq. \eqref{eq:rarb_sgwb_skewness}. This particular shape of the skewness can be produced if and only if two elements are put together: GW and Gaussianity \cite{Bernardo:2024uiq}. When pulsar red noise is considered, it turns out that the signal would again be dominated by noise, except in this case, the frequency dependence of the power spectrum, ${\cal P}(f)$, factors in and favors a resolution of the two-point signatures at higher frequency bins \cite{Bernardo:2024uiq}. We note that the two-point signatures look like Figure \ref{fig:pulsar_2p} or Eqs. (\ref{eq:rarb_sgwb_mean}-\ref{eq:rarb_sgwb_kurtosis}) in all frequency bins, only that the overall magnitudes are rescaled accordingly, i.e., if ${\tilde a}^2$ is the magnitude of the mean in the first frequency bin, $f_1\sim 1/T$, then at other frequencies, the corresponding magnitude will be ${\tilde a}^2 {\varrho}(f)$ where ${\varrho}(f)$ is a linear functional of the SGWB power spectrum ${\cal P}(f)$.

\subsection{Subluminal GWs and Non-Einsteinian Polarizations}
\label{subsec:subluminal_gws}

The spatial correlation piece/ORF of the timing residual cross correlation generalizes naturally for subluminal GWs and non-Einsteinian GW polarizations. The steps are as straightforward as Section \ref{sec:hellings_and_downs} for tensor modes in GR but with more terms accounting for non-GR polarizations, recognizable in alternative theories of gravity \cite{Chamberlin:2011ev, Clifton:2011jh, Joyce:2014kja, Nojiri:2017ncd, Kase:2018aps, Ferreira:2019xrr}. A full derivation is given in Refs. \citen{Bernardo:2022rif, Bernardo:2022xzl}, among other sources including Refs. \citen{Cordes:2024oem, Domenech:2024pow, Liang:2023ary}.

In a nutshell, the shape of gravity in a PTA can be associated with an angular power spectrum, $C_l(fD_a, fD_b)$'s such as Eq. \eqref{eq:cls_tensor} in GR \cite{Gair:2014rwa, Ng:2021waj, Bernardo:2023pwt}. This beautiful formalism enables all the past decades' work on the SGWB correlations to be put together compactly in two lines and a table:
\begin{equation}
\label{eq:powerspectrummultipoles}
    C_l^{\rm A}(fD_a, fD_b) = \dfrac{ {\cal F}_l^{\rm A}\left( fD_a,{v_{\rm g}} \right) {\cal F}_l^{\rm A}\left( fD_b,{v_{\rm g}} \right)^* }{ \sqrt{\pi} }
\end{equation}
and
\begin{equation}
\label{eq:projectionfactors}
    \dfrac{{\cal F}_l^{\rm A}\left( y,{v_{\rm g}} \right)}{N_l^{\rm A}} = \int_0^{\frac{2\pi y}{v_{\rm ph}}} dx \ v_{\rm ph} e^{ixv_{\rm ph}} \dfrac{d^q}{dx^q} \left( \dfrac{j_l(x) + r v_{\rm ph}^{-2} j_l''(x)}{x^p} \right) \,,
\end{equation}
where $v_{\rm ph}$ is the phase velocity, {$v_{\rm g}$} is the GW speed {(group velocity)}, $f$ is the GW frequency, $D_a, D_b$ are pulsar distances, and the coefficients $N_l^{\rm A}$ and indices $p, q, r$ are given in Table \ref{tab:powerspectrumcoefs} \cite{Bernardo:2023pwt}.

\begin{table}[h!]
\tbl{The SGWB angular power spectrum coefficients and indices (Eqs. \eqref{eq:powerspectrummultipoles} and \eqref{eq:projectionfactors}) for arbitrary GW modes A.}
{
\renewcommand{\arraystretch}{1.5}
\begin{tabular}{|r|r|r|r|r|}
\hline
GW modes A & $N_l^{\rm A}/\left(2 \pi i^l\right)$ & \phantom{g} $p$ & \phantom{g} $q$ & \phantom{g} $r$ \\
\hline 
Tensor & \phantom{g} $\sqrt{(l+2)!/(l-2)!}/\sqrt{2}$ & 2 & 0 & 0 \\ \hline
Vector & $\sqrt{2} \sqrt{l(l + 1)}$ & 1 & 1 & 0 \\ \hline
Scalar & $1$ & 0 & 0 & 1 \\ \hline
\end{tabular}
\label{tab:powerspectrumcoefs}
}
\end{table}

Note that theories generally have their own dispersion relation, $v_{\rm g} \equiv v_{\rm g}\left(v_{\rm ph}\right)$, for propagating degrees of freedom \cite{Liang:2023ary}. In GR, massive gravity, and the scalar modes in Horndeski theory ($f(R)$ \cite{Qin:2020hfy} and Galileon \cite{Bernardo:2022vlj}), a standard massive dispersion relation, $\omega^2 = k^2 + m^2$, suffices to turn in $v_{\rm g} = 1/v_{\rm ph}$. The ${\cal F}^{\rm A}_l(y, v{_{\rm g}})$'s are a generalization of the projection factor with finite/astrophysical pulsar distances \cite{Qin:2018yhy, Qin:2020hfy}. For scalar GWs, it is worth noting that the integration reduces to a boundary term for $l \geq 2$ that vanishes as $fD\rightarrow \infty$ \cite{Qin:2020hfy, Bernardo:2022vlj}. This shows that scalar GWs are distinguished by pronounced monopolar and dipolar contributions in the angular power spectrum. On the other hand, tensor and vector GWs are noteworthy for their significant quadrupolar and dipolar powers, respectively.

\begin{figure}[h!]
    \centering
    \includegraphics[width=0.975\linewidth]{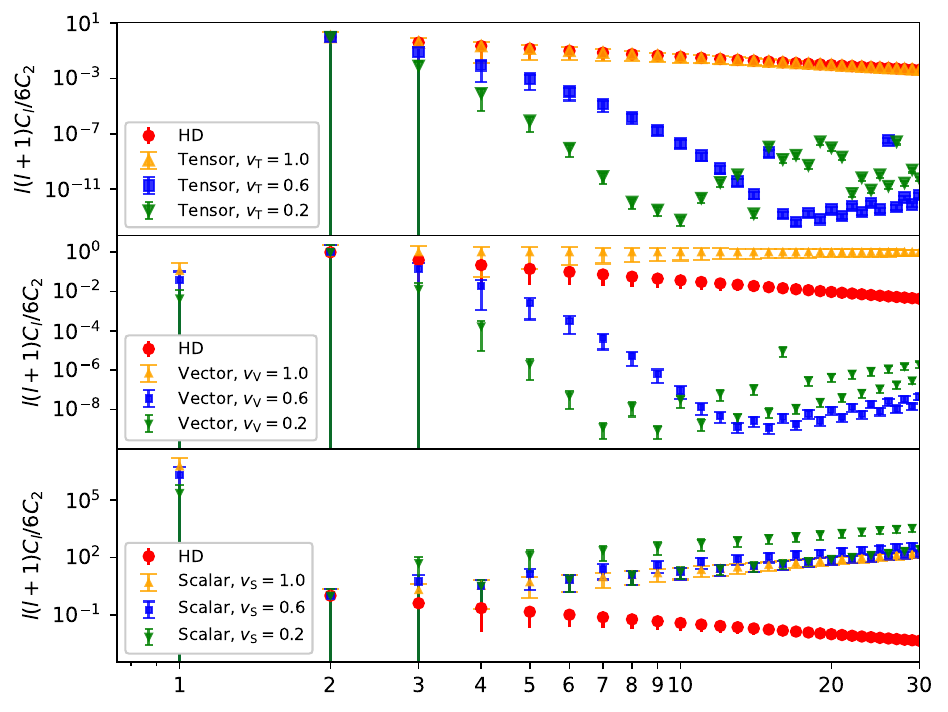}
    \caption{The SGWB angular power spectrum for tensor (top), vector (middle), and scalar (bottom) GW degrees of freedom at luminal and subluminal group speeds. For tensor, $C_0=C_1=0$; for vector, $C_0=0$; for scalar, the monopole, $C_0 \neq 0$, is excluded in the plot, with $C_0/C_1\sim 9, 25, 225$ for $v_{\rm S}\sim 1.0, 0.6, 0.2$, respectively. Red circles and error bars (2-sigma) are the standard HD correlation and its cosmic variance.}
    \label{fig:cls_gravity}
\end{figure}

The correlation angular power spectrum given by (\ref{eq:powerspectrummultipoles}-\ref{eq:projectionfactors}) and Table \ref{tab:powerspectrumcoefs} can be substituted into Eqs. \eqref{eq:gamma_harmonic_series}, \eqref{eq:cv_ps}, and \eqref{eq:variance_cls} to get the relevant ORF, cosmic variance, and the variance in the multipoles for GW modes with subluminal speeds. With $fD\sim 2000$, the results are shown in Figure \ref{fig:cls_gravity} with subluminal group speeds $v\sim 1, 1/6, 1/5$ for tensor, vector, and scalar GW degrees of freedom.

An important message of the Figure \ref{fig:cls_gravity} is that cosmic variance-limited measurements of the correlation can remarkably constrain GW propagation and non-Einsteinian GW polarizations \cite{Bernardo:2023pwt}. The distinction between different modes and GW speeds is visible even up to the quintupole ($l=5$). Metrics (motivated by a least squared-like sum and multipolar ratios) for utilizing cosmic variance-limited measurements have been discussed in Ref. \citen{Bernardo:2023pwt}, positioning PTAs to constrain nanohertz gravity in the forthcoming precision era.

The results echo that of Refs. \citen{Qin:2018yhy, Qin:2020hfy} which derived the angular power spectrum in PTA (and astrometry) at subluminal GW speeds, and complements the picture further with cosmic variance and finite distance effects \cite{Bernardo:2022rif, Bernardo:2022xzl, Bernardo:2023pwt}. For subluminal tensor and vector cases, the power at every other multipole can be seen to significantly drop relative to the dominant quadrupole. In this case, the ORF becomes approximately quadrupolar, $\gamma( \hat{e}_a \cdot \hat{e}_b )\sim P_2( \hat{e}_a \cdot \hat{e}_b )$ where $P_2(x)=(3x^2-1)/2$ \cite{Bernardo:2023pwt}. Finite distance effects characterized by the increasing power at higher multipoles can also be seen to settle in earlier in subluminal cases \cite{Bernardo:2022rif}. For subluminal scalar cases, as mentioned earlier, the correlation angular power spectrum is completely dominated by a monopole and a dipole. This remains true with finite distance effects, albeit now the higher multipoles are finite but still negligible. An important observation that can come into play to test nanohertz gravity is that the ratios of the multipoles relative to the quadrupole are tailored to specific values of the speed \cite{Bernardo:2023pwt}. Measurements of the correlation angular power spectrum up to $l=5$ \cite{NANOGrav:2023gor} in fact turn out to already be comparable to cosmic variance precision, i.e., error bars are roughly as wide as the cosmic variance, and forecasts expect that this is only going to get better \cite{Nay:2023pwu, Babak:2024yhu}. These early measurements have also hinted at a stronger quadrupolar contribution compared to the HD correlation \cite{NANOGrav:2023gor, Bernardo:2023pwt}. While it can of course be premature to take this as an observation of subluminal GWs at the nanohertz regime, it is always possible to use the present data to constrain theories, advancing knowledge \cite{Bernardo:2022vlj, Bernardo:2023mxc, Bernardo:2023bqx, Bernardo:2023zna, Wang:2023div, Wu:2023rib, Bi:2023ewq, Chen:2023uiz, Choi:2023tun, Chen:2024fir}.

We end with an illustrative calculation to show why scalar GW polarizations with a massive dispersion relation will be completely dominated by a monopole and a dipole, following simply tenets of wave superposition and partial integration. Using a massive gravity dispersion relation, it can be shown that the `scalar transverse' (ST) and `scalar longitudinal' (SL) modes in a GW superpose as \cite{Bernardo:2022vlj, Bernardo:2023pwt}
\begin{equation}
    h_{ij} \propto \varepsilon_{ij}^{\text{ST}} + \varepsilon_{ij}^{\text{SL}} \left(1-v_{\rm S}^2\right)/\sqrt{2} \,,
\end{equation}
where $v_{\rm S}$ is the group speed of scalar modes, $\varepsilon_{ij}^{\text{ST}}$ and $\varepsilon_{ij}^{\text{SL}}$ are GW polarization basis tensors analogous to the $+$ and $\times$ modes in GR. This can also be shown by linearizing a general scalar-tensor action. Thus, the scalar GWs' projection factor can be written as
\begin{equation}
    {\cal F}^{\rm S}_l \left(fD, v_{\rm S} \right) \propto - \int_0^{2\pi fDv_{\rm S}} \dfrac{dx}{v} e^{ix/v_{\rm S}} \left( R^{\rm ST}_l(x) + \dfrac{1-v_{\rm S}^2}{\sqrt{2}} R^{\rm SL}_l(x)  \right) \,,
\end{equation}
where the $R_l(x)$'s can be found in Refs. \citen{Qin:2020hfy, Bernardo:2022rif}, i.e., $R_l^\text{SL}(x) = j_l''(x)$ and $R_l^\text{ST}(x) = -\left( R_l^\text{SL}(x) + j_l(x) \right) / \sqrt{2}$. Notably, when the scalar modes propagate at the speed of light, $v_{\rm S}=1$, the SL contribution is suppressed; otherwise, the total is a mixture of SL and ST parts. Simplifying the last expression, we obtain
\begin{equation}
\begin{split}
    {\cal F}^{\rm S}_l \left(fD, v_{\rm S}\right) \propto & - \int_0^{2\pi fDv_{\rm S}} \dfrac{dx}{v_{\rm S}} e^{ix/v_{\rm S}} \left( -\dfrac{j_l''(x) + j_l(x)}{\sqrt{2}} + \dfrac{1-v_{\rm S}^2}{\sqrt{2}} j_l''(x)  \right) \\
    \propto & - \int_0^{2\pi fDv_{\rm S}} \dfrac{dx}{v_{\rm S}}  \dfrac{e^{ix/v_{\rm S}}}{\sqrt{2}} \left( -j_l(x) - v_{\rm S}^2 j_l''(x)  \right) \,.
\end{split}
\end{equation}
The calculation underlines a destructive interference in the quadrupole, octupole, and higher multipole contributions. To reveal this, we recast the integral into 
\begin{equation}
    {\cal F}^{\rm S}_l \left(fD, v_{\rm S}\right) \propto - \dfrac{1}{\sqrt{2}} \int_0^{2\pi fDv_{\rm S}} \dfrac{dx}{v_{\rm S}} \left[ \dfrac{d}{dx} \left( iv_{\rm S} e^{ix/v_{\rm S}} j_l(x) \right) - \dfrac{d}{dx} \left( v_{\rm S}^2 e^{ix/v_{\rm S}} j_l'(x) \right) \right] \,,
\end{equation}
where we have performed integration by parts with
\begin{equation}
    e^{ix/v_{\rm S}} j_l''(x) = \dfrac{d}{dx}\left(e^{ix/v_{\rm S}} j_l'(x)\right) - \dfrac{d}{dx}\left( \dfrac{i}{v_{\rm S}}e^{ix/v_{\rm S}}j_l(x) \right) - \dfrac{e^{ix/v_{\rm S}}}{v_{\rm S}^2} j_l(x) \,.
\end{equation}
For $l \geq 2$, the lower limits of the integral vanishes. The final result is
\begin{equation}
    {\cal F}^{\rm S}_l \left(fD, v_{\rm S}\right) \propto - \dfrac{e^{2\pi ifD}}{\sqrt{2}v_{\rm S}} \left[ iv_{\rm S} j_l(2\pi f Dv_{\rm S})  -  v_{\rm S}^2 j_l'(2\pi fDv_{\rm S})  \right] \,.
\end{equation}
This vanishes with $fD \rightarrow \infty$, recovering the result of Ref. \citen{Qin:2020hfy}. For astrophysical pulsar distances, the boundary terms for $l \geq 2$ remain finite, but subdominant compared to the monopole and the dipole.

\subsection{Anisotropy and Polarization}
\label{subsec:anisotropy}

It turns out that it is possible to further generalize the SGWB signal in a PTA to accommodate anisotropy and polarization components. This is discussed in a handful of trailblaizing papers on the subject matter \cite{Mingarelli:2013dsa, Taylor:2013esa, Gair:2014rwa, Kato:2015bye, Belgacem:2020nda, Chu:2021krj, Sato-Polito:2021efu, Liu:2022skj, NANOGrav:2023tcn, Tasinato:2023zcg, AnilKumar:2023yfw, Bernardo:2023jhs, AnilKumar:2023hza, Cruz:2024svc, Cruz:2024esk, Inomata:2024kzr}. This can be done conveniently by writing the power spectrum as a function of both the GW frequency and the direction, ${\cal P} \equiv {\cal P}(f, \hat{k})$, and then expressing it in terms of Stokes parameters. Below, we summarize the ORFs, obtained in the computational frame, for an anisotropic polarized SGWB in GR and beyond, with subluminal speeds and finite pulsar distances, as summarized in the most sophisticated generalization in Ref. \citen{Bernardo:2023jhs}. The computations can be done with \texttt{PTAfast}\footnote{\href{https://github.com/reggiebernardo/PTAfast}{https://github.com/reggiebernardo/PTAfast}} with a few lines of code, paving the road for constraining anisotropy and polarization in the SGWB in future studies.

For tensor GWs with frequency $f$, the ORFs for a pulsar pair $(a,b)$ with distances $D_a, D_b$, respectively, are given by \cite{Bernardo:2023jhs}
\begin{equation}
\label{eq:gammaIV_tensor_summary}
\begin{split}
    \gamma_{lm}^{I,V}\left( f D_a, fD_b, \hat{e}_a\cdot\hat{e}_b \right) = \sum_{l_1 l_2}
    & (-1)^m \left( \dfrac{2 l_1 + 1}{4\pi} \right) \left[ 1 \pm (-1)^{l + l_1 + l_2} \right] \\
    & \times C_{l_1 l_2}^{\rm \bf T}(fD_a, fD_b) Y_{l_2 m} \left( \hat{e}_a\cdot\hat{e}_b, 0 \right) \\
    & \times \sqrt{(2l+1)(2l_2 + 1)}
    \left( \begin{array}{ccc}
    l & l_1 & l_2 \\
    0 & -2 & 2
    \end{array} \right)
    \left( \begin{array}{ccc}
    l & l_1 & l_2 \\
    m & 0 & -m
    \end{array} \right) \,,
\end{split}
\end{equation}
and
\begin{equation}
\label{eq:gammaQU_tensor_summary}
\begin{split}
    \gamma_{lm}^{Q \pm i U}\left( f D_a, fD_b, \hat{e}_a\cdot\hat{e}_b \right) = \sum_{l_1 l_2}
    & (-1)^m \left( \dfrac{2l_1+1}{4\pi} \right) C_{l_1 l_2}^{\rm \bf T}(fD_a, fD_b) Y_{l_2 m}\left( \hat{e}_a\cdot\hat{e}_b, 0 \right) \\
    & \times \sqrt{ (2l+1)(2l_2 + 1) }
    \left( \begin{array}{ccc}
    l & l_1 & l_2 \\
    \mp 4 & \pm 2 & \pm 2
    \end{array} \right)
    \left( \begin{array}{ccc}
    l & l_1 & l_2 \\
    m & 0 & -m
    \end{array} \right) \,,
\end{split}
\end{equation}
where the superscripts $I, V, Q, U$ stand for Stokes parameters, i.e., an isotropic unpolarized SGWB would have $I \neq 0$ and $V = Q = U = 0$, or in terms of the correlation components, $\gamma_{lm}^I(f D_a, fD_b, \hat{e}_a\cdot\hat{e}_b)=0$ for $l \geq 1$ and $\gamma_{lm}^V(f D_a, fD_b, \hat{e}_a\cdot\hat{e}_b) = \gamma_{lm}^{Q \pm i U}(f D_a, fD_b, \hat{e}_a\cdot\hat{e}_b) = 0$ for all $l$. The $\left(\begin{array}{ccc} a & b & c \\ d & e & f \end{array}\right)$'s are Wigner-3j symbols. The $C_{l_1 l_2}^{\rm \bf T}(fD_a, fD_b)$'s are a generalization of the correlation angular power spectrum \cite{Gair:2014rwa, Qin:2018yhy, Qin:2020hfy, Ng:2021waj, Liu:2022skj, Bernardo:2022rif}, defined as
\begin{equation}
\label{eq:Cl_T_summary}
    C_{l_1 l_2}^{\rm \bf T}(fD_a, fD_b) = \dfrac{{\cal F}^{\rm \bf T}_{l_1}\left( fD_a, v \right) {\cal F}_{l_2}^{{\rm \bf T}*}\left( f D_b, v \right)}{\sqrt{\pi}}
\end{equation}
where the ${\cal F}_l^{\rm \bf T}(x, v)$'s are the projection factors\cite{Qin:2018yhy, Qin:2020hfy, Bernardo:2022xzl} given in Section \ref{subsec:subluminal_gws}. The isotropic component ($l = m = 0$) gives the HD curve in luminal GW limit, $v \sim 1$, and large pulsar distances, $fD_a \sim fD_b \sim \infty$.

For vector GWs, the correlation components are given by \cite{Bernardo:2023jhs}
\begin{equation}
\label{eq:gammaIV_vector_summary}
\begin{split}
    \gamma_{lm}^{I,V}\left( f D_a, fD_b, \hat{e}_a\cdot\hat{e}_b \right) = \sum_{l_1 l_2}
    & (-1)^{1 + m} \left( \dfrac{2l_1 + 1}{4\pi} \right) \left[ 1 \pm (-1)^{l + l_1 + l_2} \right] \\
    & \times C_{l_1 l_2}^{\rm \bf V}(fD_a, fD_b) Y_{l_2 m}\left( \hat{e}_a\cdot\hat{e}_b, 0 \right) \\
    & \sqrt{ (2l+1)(2l_2 + 1) }
    \left( \begin{array}{ccc}
    l & l_1 & l_2 \\
    0 & -1 & 1
    \end{array} \right)
    \left( \begin{array}{ccc}
    l & l_1 & l_2 \\
    m & 0 & -m
    \end{array} \right)
\end{split}
\end{equation}
and
\begin{equation}
\label{eq:gammaQU_vector_summary}
\begin{split}
    \gamma_{lm}^{Q \pm i U}\left( f D_a, fD_b, \hat{e}_a\cdot\hat{e}_b \right) = \sum_{l_1 l_2}
    & (-1)^m \left( \dfrac{2l_1 + 1}{4\pi} \right) C_{l_1 l_2}^{\rm \bf V}(fD_a, fD_b) Y_{l_2 m}\left( \hat{e}_a\cdot\hat{e}_b, 0 \right)  \\
    & \sqrt{ (2l+1)(2l_2 + 1) }
    \left( \begin{array}{ccc}
    l & l_1 & l_2 \\
    \mp 2 & \pm 1 & \pm 1
    \end{array} \right)
    \left( \begin{array}{ccc}
    l & l_1 & l_2 \\
    m & 0 & -m
    \end{array} \right) \,,
\end{split}
\end{equation}
where the $C_{l_1 l_2}^{\rm \bf V}(fD_a, fD_b)$'s are the vector SGWB correlation angular power spectrum,
\begin{equation}
\label{eq:Cl_V_summary}
    C_{l_1 l_2}^{\rm \bf V}(fD_a, fD_b) = \dfrac{{\cal F}^{\rm \bf V}_{l_1}\left( fD_a, v \right) {\cal F}_{l_2}^{{\rm \bf V}*}\left( f D_b, v \right)}{\sqrt{\pi}} \,.
\end{equation}
Note that the Wigner-3j symbols are indicative of the spin of the underlying field, distinguishing the tensor and vector cases.

For scalar GWs, there is no need to breakdown the correlation in Stokes parameters. In this case, the correlation components can be easily written as \cite{Bernardo:2023jhs}
\begin{equation}
\label{eq:gamma_scalar_summary}
\begin{split}
    \gamma_{lm} \left( fD_a, fD_b, \hat{e}_a\cdot\hat{e}_b \right) = \sum_{l_1 l_2}
    & (-1)^m \left( \dfrac{2l_1 + 1}{4\pi} \right) C_{l_1 l_2}^{\rm \bf S}(fD_a, fD_b) Y_{l_2 m}\left( \hat{e}_a\cdot\hat{e}_b, 0 \right) \\
    & \times \sqrt{ (2l+1)(2l_2 + 1) }
    \left( \begin{array}{ccc}
    l & l_1 & l_2 \\
    0 & 0 & 0
    \end{array} \right)
    \left( \begin{array}{ccc}
    l & l_1 & l_2 \\
    m & 0 & -m
    \end{array} \right) \,,
\end{split}
\end{equation}
where $C_{l_1 l_2}^{\rm \bf S}(fD_a, fD_b)$'s are the scalar SGWB correlation angular power spectrum
\begin{equation}
    C_{l_1 l_2}^{\rm \bf S}(fD_a, fD_b) = 32 \pi^2 {\cal F}^{\rm \bf S}_{l_1}\left( fD_a, v \right) {\cal F}_{l_2}^{{\rm \bf S} *}\left( f D_b, v \right) \,.
\end{equation}
The Stokes parameter decomposition is not utilized since scalar modes carry a longitudinal (SL) polarization, 
stretching test masses in the direction of propagation of a GW, and a transverse breathing mode (ST), 
stretching space equally in all directions transverse to the direction of propagation.
As with tensor and vector cases, the Wigner-3j symbol gives away the spin-$0$ nature of a scalar field.

We emphasize that all of the expressions above in this section properly reduce to the ones written previously in the isotropic limit.

\section{Outlook}
\label{sec:outlook}

The compelling evidence of the nanohertz stochastic gravitational wave background detected by the pulsar timing arrays should be regarded as no less than an astronomical milestone. Beyond the science, this achievement is marked by a great deal of resilience---of dedicated generations of individuals and communities working tirelessly to keep the telescopes operational, ensuring the collection of the precise data that has led to this breakthrough.

Such commitment to the craft could only be matched by a robust theoretical foundation, which the field has received since the pioneering works of Sazhin \cite{Sazhin:1978myk}, Detweiler \cite{Detweiler:1979wn}, Hellings-Downs \cite{Hellings:1983fr}, and Blandford-Narayan-Romani \cite{1984JApA....5..369B} in the 1980s, continuing through the influential contributions of Phinney \cite{Phinney:2001di}, Wyithe-Loeb \cite{Wyithe:2002ep} and Sesana-Haardt-Madau-Volonteri \cite{Sesana:2004sp} in the 2000s. This early momentum has clearly evolved into a significant and flourishing domain during the gravitational wave era of the 2010s and 2020s (see our list of milestones in the Introduction and the references therein). The theoretical framework developed in this review has provided an adamantium skeleton for stochastic gravitational wave background detection, offering precise and accurately derived expressions for the signal and the correlation in a PTA (Sections \ref{sec:pulsar_timing_model}-\ref{sec:hellings_and_downs}). Furthermore, we have explored aspects of the signal that could potentially enhance detection, such as cosmic variance (Section \ref{subsec:cosmic_variance}), and have introduced new observational milestones to constrain theories, including the cumulants of the one- and two-point functions (Section \ref{subsec:cumulants}). Additionally, we have highlighted how pulsar timing arrays could constrain non-Einsteinian polarizations, subluminal gravitational waves (Section \ref{subsec:subluminal_gws}), and anisotropy and polarization (Section \ref{subsec:anisotropy}).

One of the most intriguing debates in the field today centers on the origin of the detected signal---whether it arises from astrophysical sources, such as supermassive black hole binaries, or from a cosmological background, potentially linked to phenomena like primordial gravitational waves or phase transitions in the early Universe. This debate underscores the richness of the field and the potential for future discoveries that could reshape our understanding of the Universe.

Looking forward, the future of this field is exceptionally bright \cite{Siemens:2013zla, Moore:2014eua, Vigeland:2016nmm, Hazboun:2019vhv}, looking toward source statistics \cite{Allen:2024rqk, Allen:2024mtn}, resolving individual binaries \cite{Mingarelli:2012hh, Susobhanan:2020iyj, Susobhanan:2022nzv, EPTA:2023gyr, NANOGrav:2023wsz, NANOGrav:2023pdq}, with significant advancements expected in both the theoretical \cite{OBeirne:2019lwp} and instrumental arenas \cite{Lazio:2013mea, Weltman:2018zrl, ChandraJoshi:2022etw}, as well as in data processing and analysis techniques. Continued collaboration within the scientific community will be crucial to unraveling the mysteries of the nanohertz stochastic gravitational wave background, as we push the boundaries of our knowledge and refine detection capabilities.

\section*{Acknowledgments}
We are grateful to Marc Kamionkowski, Zu-Cheng Chen, Meng-Xiang Lin, and Qiuyue Liang for feedback on a preliminary draft. We thank our esteemed colleagues Guo-Chin Liu, Stephen Appleby, Achamveedu Gopakumar, Subhajit Dandapat, and Debabrata Deb whom we have exchanged plenty of exciting discussions with. RCB is supported by an appointment to the JRG Program at the APCTP through the Science and Technology Promotion Fund and Lottery Fund of the Korean Government, and was also supported by the Korean Local Governments in Gyeongsangbuk-do Province and Pohang City. This work was supported in part by the National Science and Technology Council of Taiwan, Republic of China, under Grant No. NSTC 113-2112-M-001-033.

\appendix


\end{document}